\documentclass[conference]{IEEEtran}

\usepackage[table]{xcolor}
\usepackage{algorithm}
\usepackage{algpseudocode}
\usepackage{subfigure}
\usepackage{graphicx}
\usepackage{amsmath}
\usepackage{relsize}

\let\labelindent\relax
\usepackage{enumitem}

\newcommand{\minus}{\scalebox{0.75}[1.0]{$-$}}
\newcommand{\plus}{\scalebox{0.9}[0.9]{$+$}}

\newcommand{\VMusagetrim}[2][]{%
    \includegraphics[scale=0.27,trim = 240 240 300 120,clip]%
        {#2}%
}
\newcommand{\RTtrim}[2][]{
    \includegraphics[scale=0.50,trim = 30 190 480 110,clip]%
        {#2}%
}

\begin{document}

\title{A Comparison of Reinforcement Learning Techniques for Fuzzy Cloud Auto-Scaling}

\author{
  \IEEEauthorblockN{
    Hamid Arabnejad\IEEEauthorrefmark{1}, 
    Claus Pahl\IEEEauthorrefmark{2}, 
    Pooyan Jamshidi\IEEEauthorrefmark{3} and
    Giovani Estrada\IEEEauthorrefmark{4}
    }
  \IEEEauthorblockA{
    \IEEEauthorrefmark{1}IC4, Dublin City University, Dublin, Ireland }
    \IEEEauthorblockA{\IEEEauthorrefmark{2}Free University of Bozen-Bolzano, Bolzano, Italy}
    \IEEEauthorblockA{\IEEEauthorrefmark{3}Imperial College London, London, UK}
    \IEEEauthorblockA{\IEEEauthorrefmark{4}Intel, Leixlip, Ireland}
    }

\maketitle

\begin{abstract}
A goal of cloud service management is to design self-adaptable auto-scaler to react to workload fluctuations and changing the resources assigned. The key problem is how and when to add/remove resources in order to meet agreed service-level agreements. Reducing application cost and guaranteeing service-level agreements (SLAs) are two critical factors of dynamic controller design. In this paper, we compare two dynamic learning strategies based on a fuzzy logic system, which learns and modifies fuzzy scaling rules at runtime. A self-adaptive fuzzy logic controller is combined with two reinforcement learning (RL) approaches: (i) Fuzzy SARSA learning (\texttt{FSL}) and (ii) Fuzzy Q-learning (\texttt{FQL}). As an \textit{off-policy} approach, Q-learning learns independent of the policy currently followed, whereas SARSA as an \textit{on-policy} always incorporates the actual agent's behavior and leads to faster learning.
Both approaches are implemented and compared in their advantages and disadvantages, here in the OpenStack cloud platform. We demonstrate that both auto-scaling approaches can handle various load traffic situations, sudden and periodic, and delivering resources on demand while reducing operating costs and preventing SLA violations. The experimental results demonstrate that \texttt{FSL} and \texttt{FQL} have acceptable performance in terms of adjusted number of virtual machine targeted to optimize SLA compliance and response time.

\end{abstract}


\IEEEpeerreviewmaketitle

\section{Introduction}
Automated elasticity and dynamism, as two important concepts of cloud computing, are beneficial for application owners. \textit{Auto-scaling} system is a process that automatically scales the number of resources and maintains an acceptable Quality of Service (QoS) \cite{lorido2014review}. However, from the perspective of the user, determining when and how to resize the application makes defining a proper auto-scaling process difficult. Threshold-based auto-scaling approaches are proposed for scaling application by monitoring metrics, but setting the corresponding threshold conditions still rests with the user. Recently, automatic decision-making approaches, such as reinforcement learning (RL) \cite{sutton1998introduction}, have become more popular. The key advantage of these methods is that \textit{prior} knowledge of the application performance model is not required, but they rather learn it as the application runs. 

Our motivation here is to compare two different auto-scaling services that will automatically and dynamically resize user application to meet QoS requirements cost-effectively. We consider extensions of two classic RL algorithms, namely SARSA and Q-Learning, for the usage with a fuzzy auto-scaling controller for dynamic resource allocations. RL is defined as interaction process between a learning agent (the auto-scaling controller) and its environment (the target could application). The main difference between SARSA and Q-learning is that SARSA compares the current state vs.\ the actual next state, whereas Q-Learning compares the current state vs.\ the best possible next states.

Generally, RL approaches suffer from the size of the table needed to store state-action values. As a solution, a fuzzy inference system offers a possible solution to reducing the state space. A fuzzy set is a mapping of real state to a set of fuzzy labels. Therefore, many states can be represented by only a few fuzzy states. Thus, we base our investigation on a fuzzy controller \cite{Jamshidi2015}.
The combination of fuzzy logic control and RL approaches results in a self-adaptive mechanism where the fuzzy logic control facilitates the reasoning at a higher level of abstraction, and the RL approaches allow to adjust the auto-scaling controller. This paper extend previous results \cite{jamshidi2014autonomic,jamshidi2016autonomic} as follows. First, we specifically focus on architecture, implementation and experimentation aspects in OpenStack. Then, we utilise SARSA approach as an \textit{on-policy} learning algorithm against Q-learning which is an \textit{off-policy} approach. The advantage of using SARSA, due to following the action which is actually being taken in the next step, is the policy that it follows will be more optimal and learning will be faster. Furthermore, a comparison between the two strategies will be provided. The comparison analysis is an important goal to know the performance and scalability of each RL approaches under different workload patterns.

The contributions of this paper are:
\begin{itemize}
	\item a review of cloud auto-scaling approaches;
	\item integrate RL and fuzzy approaches  as an automatic decision-making in a real cloud controller;
	\item implementation of Fuzzy RL approaches in OpenStack (industry-standard IaaS platform);
	\item extensive experimentation and evaluation of wide range of workload patterns;
	\item comparison between two RL approaches, SARSA and Q-learning in terms of quality of results
\end{itemize}
We show that the auto-scaling approaches can handle various load traffic situations, delivering resources on demand while reducing infrastructure and management costs alongside the comparison between both proposed approaches. The experimental results show promising performance in terms of resource adjustment to optimize Service Level Agreement (SLA) compliance and response time while reducing provider costs.

The paper is organized as follows. Section \ref{sec_2} describes auto-scaling process briefly, and discusses on related research in this area, Section \ref{sec_3} describes the OpenStack architecture and orchestration, Section \ref{sec_4} describes our proposed \texttt{FSL} approach in details followed by implementation in Section \ref{sec_5}. A detailed experiment-based evaluation follows in Section \ref{sec_6}.

\section{Background and Related Work}\label{sec_2}

The aim of auto-scaling approaches is to acquire and release resources dynamically while maintaining an acceptable QoS \cite{lorido2014review}. The auto-scaling process is usually represented and implemented by a MAPE-K (Monitor, Analyze, Plan and Execute phases over a Knowledge base) control loop \cite{huebscher2008survey}.

An auto-scaler is designed with particular goal, relying on scaling abilities offered by the cloud providers or focusing on the structure of the target application. We can classify auto-scaling approaches based on usage theory and techniques:

\subsection{Threshold-based rules}

Threshold-based rules are the most popular approach offered by many platforms such as Amazon EC2\footnote{http://aws.amazon.com/ec2}, Microsoft Azure\footnote{http://azure.microsoft.com} or OpenStack\footnote{https://www.openstack.org}. Conditions and rules in threshold-based approaches can be defined based on one or more performance metrics, such as CPU load, average response time or request rate. 
Dutreilh et al.\ \cite{dutreilh2010data} investigate horizontal auto-scaling using threshold-based and reinforcement learning techniques. In \cite{han2012lightweight}, the authors describe a lightweight approach that operates fine-grained scaling at resource level in addition to the VM-level scaling in order to improve resource utilization while reducing cloud provider costs. Hasan et al.\ \cite{hasan2012integrated} extend the typical two threshold bound values and add two levels of threshold parameters in making scaling decisions. Chieu et al.\ \cite{chieu2009dynamic} propose a simple strategy for dynamic scalability of PaaS and SaaS web applications based on the number of active sessions and scaling the VMs numbers if all instances have active sessions exceed particular thresholds. 
The main advantage of threshold-based auto-scaling approaches is their simplicity which make them easy to use in cloud providers and also easy to set-up by clients. However, the performance depends on the quality of the thresholds.

\subsection{Control theory}

Control theory deals with influencing the behaviour of dynamical systems by monitoring output and comparing it with reference values. By using the feedback of the input system (difference between actual and desired output level), the controller tries to align actual output to the reference. For auto-scaling, the reference parameter, i.e., an object to be controlled, is the targeted SLA value \cite{Jamshidi2016b}. 
The system is the target platform and system output are parameters to evaluate system performance (response time or CPU load). Zhu and Agrawal \cite{zhu2010resource} present a framework using  Proportional-Integral (PI) control, combined with a reinforcement learning component in order to minimize application cost. Ali-Eldin et al.\ \cite{ali2012adaptive,ali2012efficient} propose two adaptive hybrid reactive/proactive controllers in order to support service elasticity by using the queueing theory to estimate the future load. Padala et al.\ \cite{padala2009automated} propose a feedback resource control system that automatically adapts to dynamic workload changes to satisfy service level objectives. They use an online model estimator to dynamically maintain the relationship between applications and resources, and a two-layer multi-input multi-output (MIMO) controller that allocates resources to applications dynamically. Kalyvianaki et al.\ \cite{kalyvianaki2009self} integrate a Kalman filter into feedback controllers that continuously detects CPU utilization and dynamically adjusts resource allocation in order to meet QoS objectives. 

\subsection{Time series analysis}

The aim of time series analysis is to carefully collect and study the past observations of historical collect data to generate future value for the series. Some forecasting models such as Autoregressive (AR), Moving Average (MA) and Autoregressive Moving Average (ARMA) focus on the direct prediction of future values, whereas other approach such as pattern matching and Signal processing techniques first try to identify patterns and then predict future values. Huang et al.\ \cite{huang2012resource} proposed a prediction model (for CPU and memory utilization) based on double exponential smoothing to improve the forecasting accuracy for resource provision. Mi et al. \cite{mi2010online} used Brown’s quadratic exponential smoothing to predict the future application workloads alongside of a genetic algorithm to find a near optimal reconfiguration of virtual machines. By using ARMA, Roy et al.\ \cite{roy2011efficient} presented a look-ahead resource allocation algorithm to minimizing the resource provisioning costs while guaranteeing the application QoS in the context of auto-scaling elastic clouds. By combining a sliding window approach over previous historical data and artificial neural networks (ANN), Islam et al.\ \cite{islam2012empirical} proposed adaptive approach to reduce the risk of SLA violations by initializing VMs and perform their boot process before resource demands. Gong et al.\ \cite{gong2010press} used the Fast Fourier Transform to identify repeating patterns. The Major drawback relies on this category is the uncertainty of prediction accuracy that highly on target application, input workload pattern, the selected metric, the history window and prediction interval, as well as on the specific technique being used \cite{lorido2014review}.

\subsection{Reinforcement learning (RL)}

\begin{figure}[ht!]
\centering
\vspace*{-0.1cm}
	\scalebox{0.15}{\includegraphics[trim=10 100 10 130,clip=true]{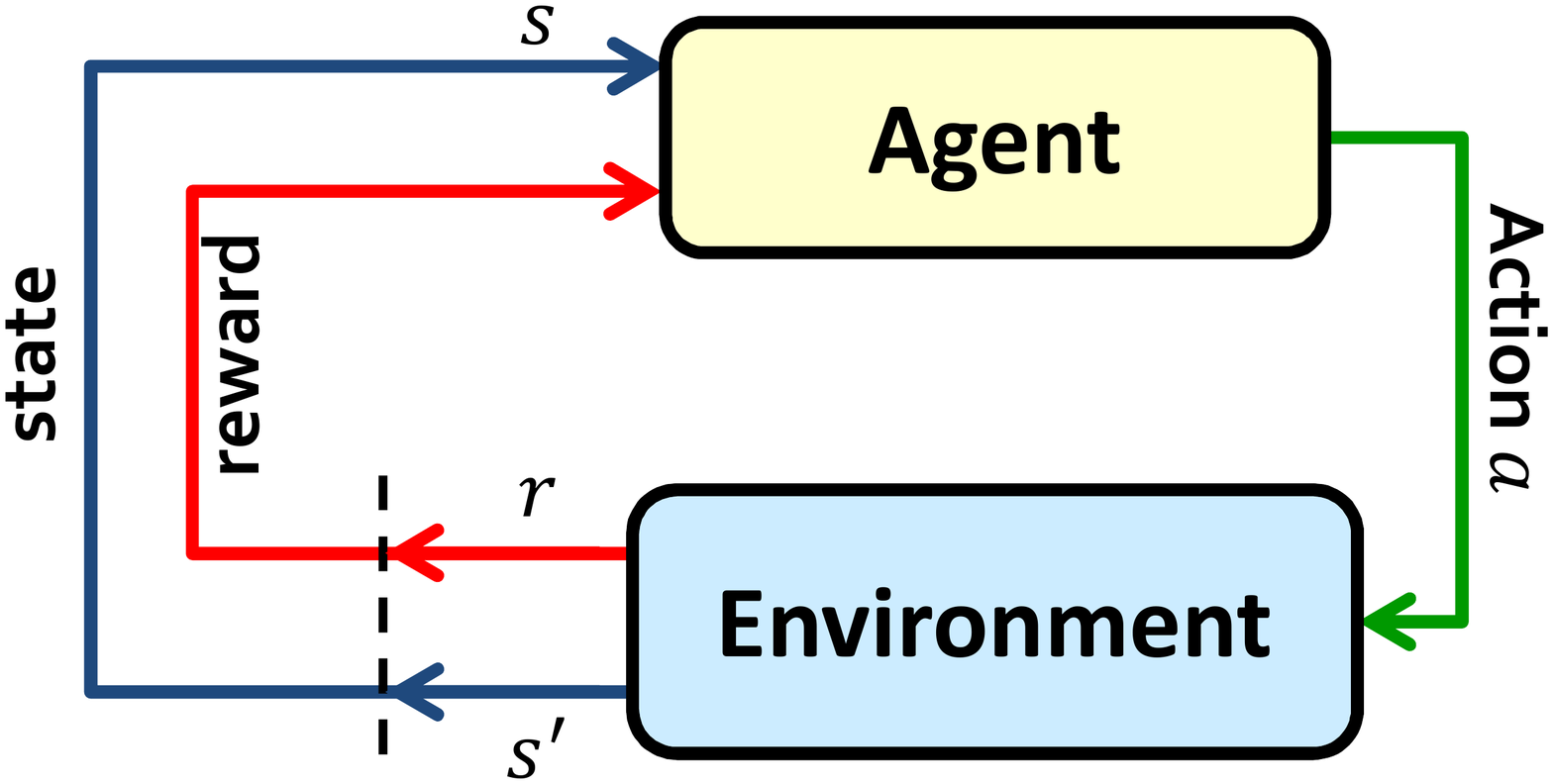}}
	\caption{The standard architecture of the RL algorithm}
	\label{RL_architecture}
\vspace*{-0.1cm}%
\end{figure}

RL \cite{sutton1998introduction} is learning process of an agent to act in order to maximize its rewards. The standard RL architecture is given in Figure \ref{RL_architecture}. The agent is defined as an auto-scaler, the action is scaling up/down, the object is the target application and the reward is the performance improvement after applying the action. The goal of RL is how to choose an action in response to a current state to maximize the reward. There are several ways to implement the learning process. Generally, RL approaches learn estimates of the Initialized Q-values $Q(s,a)$, which maps all system states $s$ to their best action $a$. We initialise all $Q(s,a)$ and during learning, choose an action $a$ for state $s$ based on $\epsilon$-greedy  policy and apply it in the target platform. Then, we observe the new state $s'$ and reward $r$ and update the Q-value of the last state-action pair $Q(s,a)$ with respect to the observed outcome state ($s'$) and reward ($r$). 

Two well-known RL approaches are SARSA and Q-learning \cite{sutton1998introduction}. Dutreilh et al.\ \cite{dutreilh2011using} use an appropriate initialization of the Q-values to obtain a good policy from the start as well as convergence speedups to quicken the learning process for short convergence times. Tesauro et al.\ \cite{tesauro2006hybrid} propose a hybrid learning system by combining queuing network model and SARSA learning approach to make resource allocation decisions based on application workload and response time.

The important feature of RL approaches is learning without prior knowledge of the target scenario and ability to online learn and update environmental knowledge by actual observations. However, there are some drawbacks in this approach such as taking long time to converge to optimal or near optimal solution for solving large real world problems and requiring good initialization of the Q-function.

\section{OpenStack orchestration}\label{sec_3}
\begin{figure}[H]
\centering
	\subfigure{\scalebox{0.25}{\includegraphics[trim=20 10 30 10,clip=true]{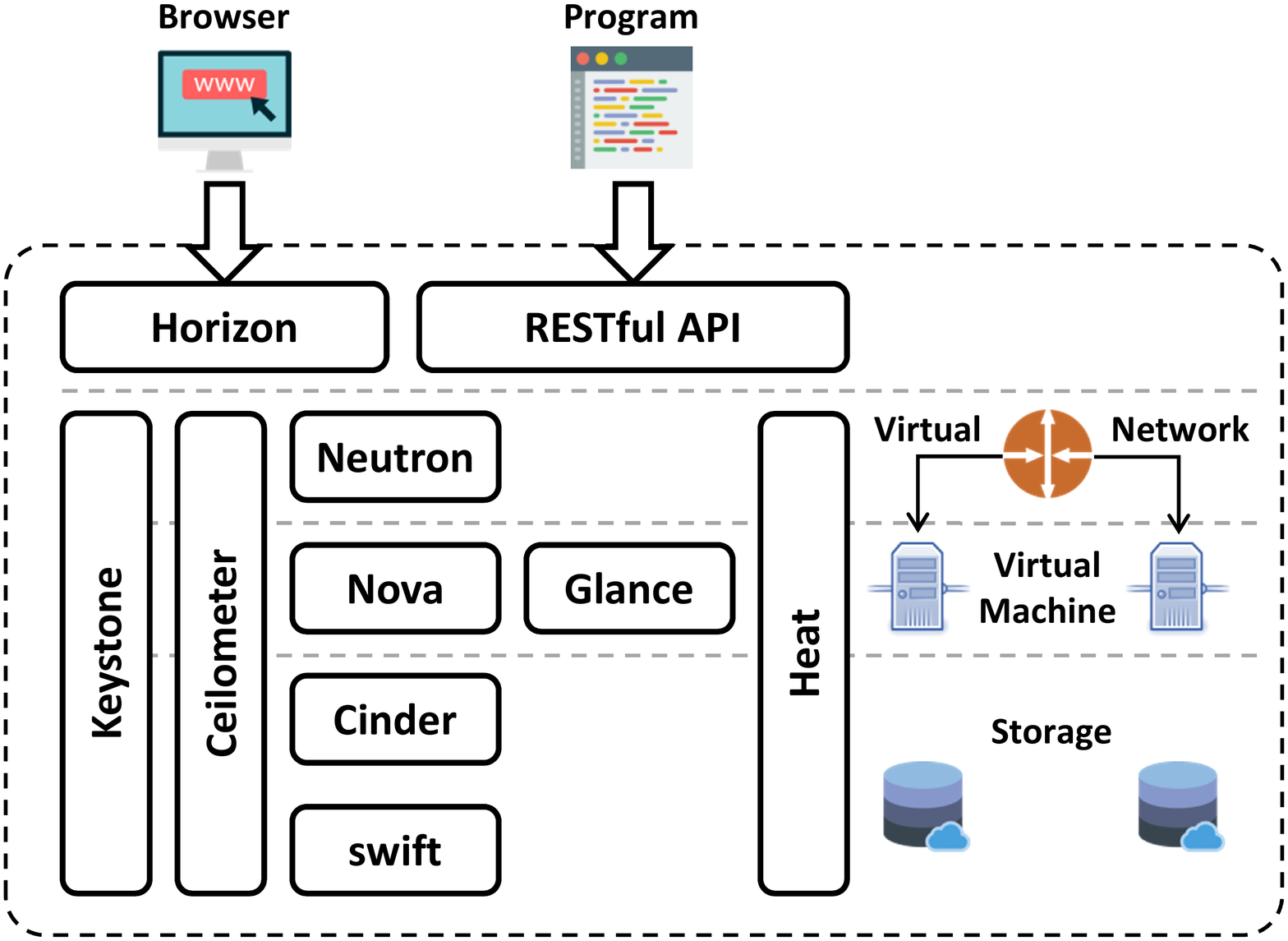}}}
	\caption{An OpenStack block diagram}
	\label{openstack_architecture}
\end{figure}

OpenStack is an IaaS open-source platform, used for building public and private clouds. It consists of interrelated components that control hardware pools of processing, storage, and networking resources throughout a data center. Users either manage it through a web-based dashboard, through command-line tools, or through a RESTful API. Figure \ref{openstack_architecture} shows a high-level overview of OpenStack core services.

In OpenStack, 1) Neutron is a system for managing networks and IP addresses; 2) Nova is the computing engine for deploying and managing virtual machines; 3) Glance supports discovery, registration and delivery for disk and server images; 4) Ceilometer provides telemetry services to collect metering data; 5) Keystone provides user/service/endpoint authentication and authorization and 6) Heat is a service for orchestrating the infrastructure needed for cloud applications to run. 

\begin{figure}[h]
\centering
	\scalebox{0.25}{\includegraphics[trim=20 20 20 25,clip=true]{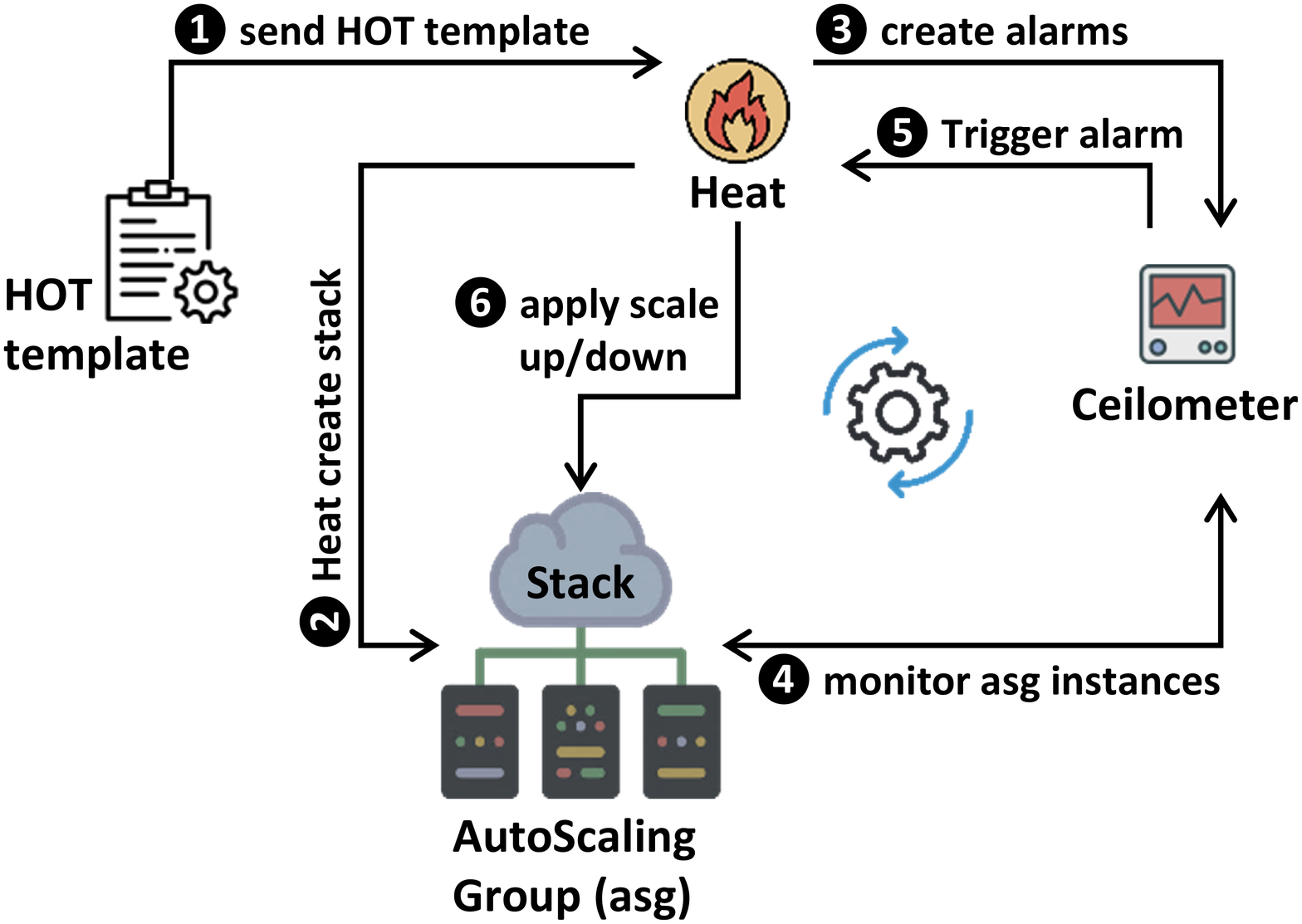}}
	\caption{Heat + Ceilometer architecture}
	\label{heat_Ceilometer}
\end{figure}

\textit{OpenStack Orchestration} is about managing the infrastructure required by a cloud application for its entire lifecycle. Orchestration automates processes which provision and integrate cloud resources such as storage, networking and instances to deliver a service defined by policies. Heat, as OpenStack's main orchestration component, implements an engine to launch multiple composite applications described in text-based templates. Heat templates are used to create stacks, which are collections of resources such as compute instance, floating IPs, volumes, security groups or users, and the relationship between these resources. Heat along with Ceilometer can create an auto-scaling service. By defining a scaling group (e.g., compute instance) alongside using monitoring alerts (such as CPU utilization) provided by Ceilometer, Heat can dynamically adjust the resource allocation, i.e., launching resources to meet application demand and removing them when no longer required, see Figure \ref{heat_Ceilometer}. Heat executes Heat Orchestration Templates (HOT), written in YAML.

By sending a HOT template file to the Heat engine, a new autoscaling group (\texttt{asg}) is created by launching a group of VM instances. The maximum and minimum number of instances should be defined in HOT file. Then, Ceilometer alarms that monitor all of the instances in \texttt{asg} are defined. Basically, at each Ceilometer interval time, the system checks the alarm metric and if it passed the defined threshold values, the scaling up/down policy will be performed based on defined action in the HOT file. During the life cycle of the application, all checking, testing and actions  are performed automatically.

\section{On-Policy and Off-Policy RL Auto-Scaling}\label{sec_4}

In \cite{jamshidi2014autonomic}, an elasticity controller based on a fuzzy logic system is proposed. The motivation factor for using fuzzy control systems is the fact that they make it easier to incorporate human knowledge in the decision-making process in the form of fuzzy rules, but also reduce the state space. 

We extend the fuzzy controller in the form of a SARSA-based Fuzzy Reinforcement Learning algorithm as an \textit{on-policy} learning approach, called \texttt{FSL}, and describe this in more detail.  Then, we related this to a Q-Learning-based \textit{off-policy} learning approach, called \texttt{FQL}, by describing the differences.

\subsection{Reinforcement Learning (RL)}

Reinforcement learning \cite{sutton1998introduction} is learning by trial and error to map situations to actions, which aims to maximize a numerical reward signal. The learning process consists of two components: a) an agent (i.e., the auto-scaler) that executes actions and observes the results and b) the environment (i.e., the application) which is the target of the actions. In this schema, the auto-scaler as an agent interacts with an environment through applying \textit{scaling actions} and receiving a response, i.e., the \textit{reward}, from the environment. Each \textit{action} is taken depending on the current \textit{state} and other environmental parameters such as the input workload or performance, which moves the agent to a different \textit{state}. According to the \textit{reward} from system about the action quality, the auto-scaler will learn the best scaling action to take through a trial-and-error.

\subsection{Fuzzy Reinforcement Learning (\texttt{FRL})}

We extend fuzzy auto-scaling with two well-known RL strategies, namely Q-learning and SARSA. We start with a brief introduction of the fuzzy logic system  and then describe proposed \texttt{FSL} and \texttt{FQL} approaches.

\begin{figure}[H]
\belowcaptionskip = -10pt
\centering
	\subfigure{\scalebox{0.30}{\includegraphics[trim=25 140 15 150,clip=true]{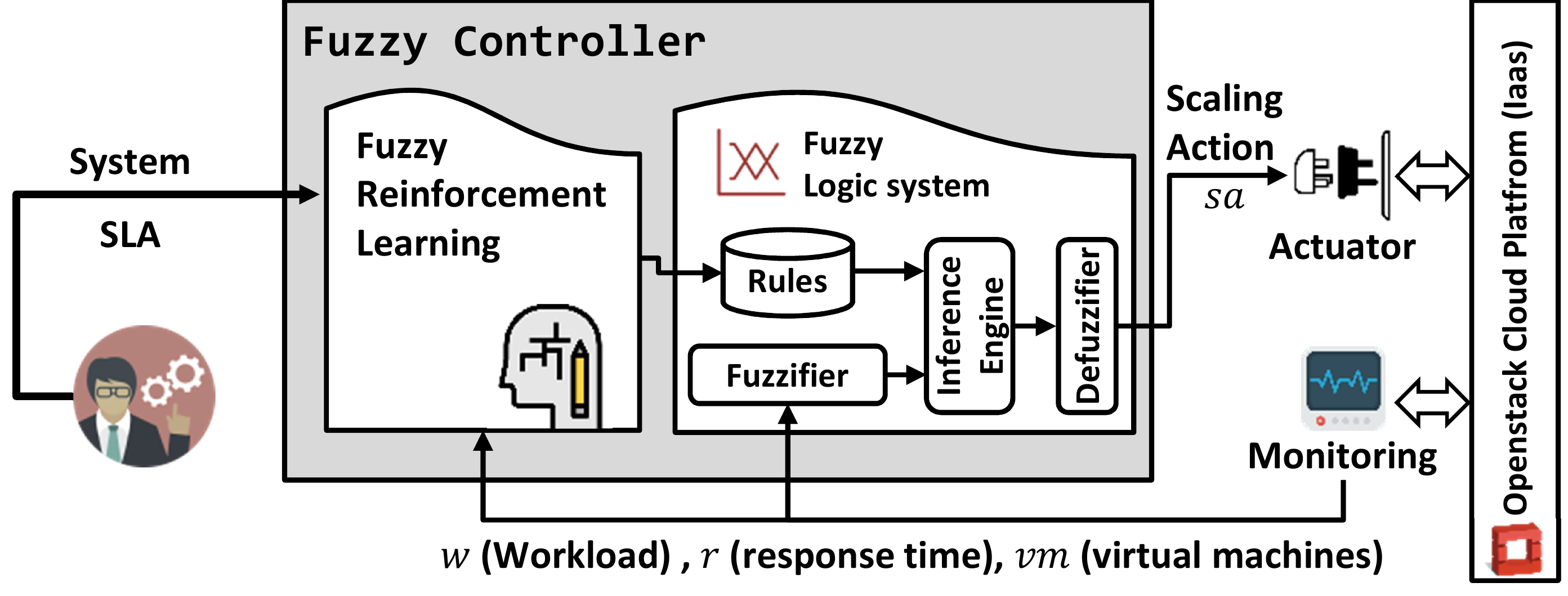}}}
	\caption{\texttt{FRL} (logical) architecture}
	\label{FRL_architecture}
\end{figure}

The purpose of the fuzzy logic system is to model a human knowledge. Fuzzy logic allows us to convert expert knowledge in the form of rules, apply it in the given situation and conclude a suitable and optimal action according to the expert knowledge. Fuzzy rules are collections of \texttt{IF-THEN} rules that represent human knowledge on how to take decisions and control a target system. 
Figure \ref{FRL_architecture} illustrates the main building blocks of a Fuzzy Reinforcement Learning (\texttt{FRL}) approach. During the lifecycle of an application, \texttt{FRL} guides resource provisioning. More precisely, \texttt{FRL} follows the autonomic MAPE-K loop by monitoring continuously different characteristics of the application (e.g., workload and response time), verifying the satisfaction of system goals and adapting the resource allocation in order to maintain goal satisfaction. The goals (i.e., SLA, cost, response time) are reflected in the reward function that we define later in this section.

The monitoring component collects required metrics such as the workload ($w$), response time ($rt$) and the number of virtual machines ($vm$) and feeds both to the controller and the knowledge learning component. The controller is a fuzzy logic controller that takes the observed data, calculates the scaling action based on  monitored input data and a set of rules, and as output returns the scaling action ($sa$) in terms of an increment/decrement in the number of virtual machines. 
The actuator issues adaptation commands from the controller at each control interval to the underlying cloud platform.

\begin{figure}[H]
\belowcaptionskip = -10pt
\centering
	\subfigure{\scalebox{0.26}{\includegraphics[trim=85 270 70 145,clip=true]{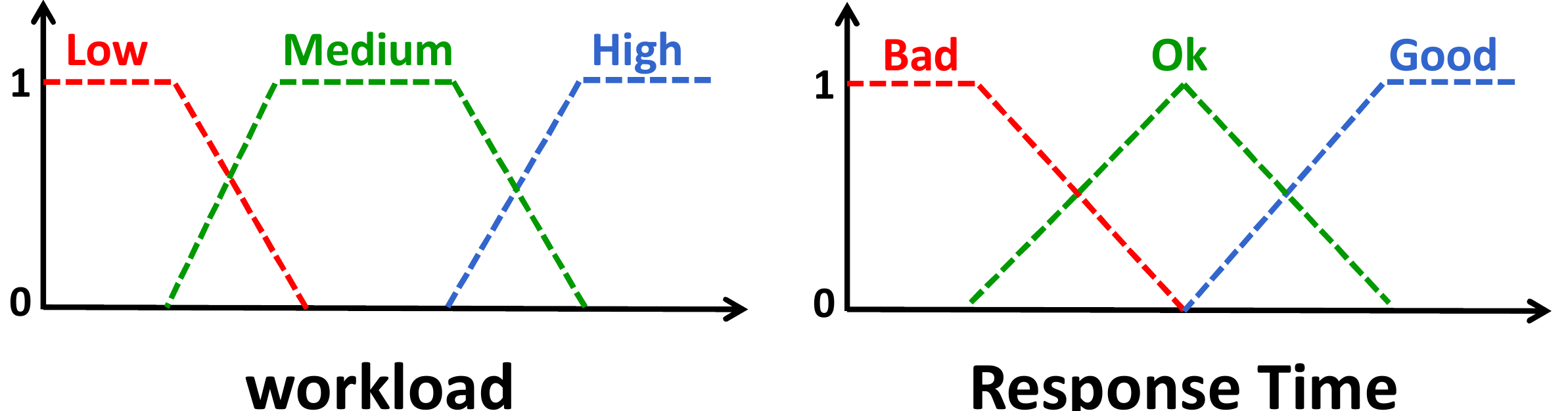}}}
	\caption{Fuzzy membership functions for auto-scaling variables}
	\label{FQL4_membership}
\end{figure}

Generally, the design of a fuzzy controller involves all parts related to membership functions, fuzzy logic operators and \texttt{IF-THEN} rules.
The first step is to partition the state space of each input variable into fuzzy sets through membership functions. The membership function, denoted by $\mu(x)$, quantifies the degree of membership of an input signal $x$ to the fuzzy set $y$. Similar to \cite{jamshidi2014autonomic}, the membership functions, depicted in Figure \ref{FQL4_membership}, are triangular and trapezoidal. Three fuzzy sets have been defined for each input (i.e., workload and response time) to achieve a reasonable granularity in the input space while keeping the number of states small.

For the inference mechanism, the elasticity policies are defined as rules: \texttt{"IF ($w$ is $high$) AND ($rt$ is $bad$) THEN ($sa+=2$)"}, where $w$ and $rt$ are monitoring metrics stated in the SLA and $sa$ is the change constant value in the number of deployed nodes, i.e., the VMs numbers.

Once the fuzzy controller is designed, the execution of the controller is comprised of three steps (cf.\ middle part of Figure \ref{FRL_architecture}): (i) fuzzification of the inputs, (ii) fuzzy reasoning, and (iii) defuzzification of the output. The fuzzifier projects the crisp data onto fuzzy information using membership functions. The fuzzy engine reasons based on information from a set of fuzzy rules and derives fuzzy actions. The defuzzifier reverts the results back to crisp mode and activates an adaptation action. This result is enacted by issuing appropriate commands to the underlying platform fabric.

Based on this background, we can now combine the fuzzy logic controller with the two RL approaches.

\subsection{Fuzzy SARSA Learning (\texttt{FSL})}

By using RL approaches instead of relying on static threshold values to increase/decrease the amount of VMs, the performance of target application can be captured after applying each $sa$ decision. In this paper, we use SARSA and Q-learning as RL approaches that we combine with the fuzzy controller. In this schema, a state $s$ is modeled by a triple ($w$,$rt$,$vm$) for which an  \textit{RL} approach looks for best action $a$ to execute. The combination of the fuzzy logic controller with SARSA \cite{sutton1998introduction} learning, called \texttt{FSL}, is explained in the following.

\begin{enumerate}[leftmargin=*]

	\item Initialize the $q$-values: unlike the  threshold policy, the \textit{RL} approach captures history information of a target application into a value table. Each member of the $q$-value table is assigned to a certain rule that describes some state-action pairs and is updated during the learning process. It can tell us the performance of taking the action by taking into account the reward value. In this study, we set all $q$-values to $0$ as simplest mode.

	\item Select an action: to learn from the system environment, we need to explore the knowledge that has already been gained. The approach is also known as the \textit{exploration/exploitation} strategy. $\epsilon$-greedy is known as a standard exploration policy \cite{sutton1998introduction}. Most of the time (with probability $1-\epsilon$), the action with the best reward will be selected or a random action will be chosen (with low probability $\epsilon$) in order to explore non-visited actions. The purpose of this strategy is to encourage exploration. After a while, by decreasing $\epsilon$, no further  exploration is made.

	\item Calculate the control action inferred by fuzzy logic controller: The fuzzy output is a weighted average of the consequences of the rule, which can be written as:
\begin{equation}
 a = {\sum\limits_{i=1}^{N}}\mu_i(x)\times a_i
 \label{output_eq}
\end{equation}
where $N$ is the number of rules, $\mu_i(x)$ is the firing degree of the rule $i$ (or the degree of truth) for the input signal $x$ and $a_i$ is the consequent function for the same rule.

	\item Approximate the $Q$-function from the current $q$-values and the firing level of the rules: In classical RL, only one state-action pair (rule) can be executed at once, which is not true for the condition of fuzziness. In a fuzzy inference system, more rules can be taken and an action is composed of these rules \cite{Fang201611}. Hence, the $Q$ value of an action $a$ for the current state $s$ is calculated by:
\begin{equation}
	Q(s,a)={\mathlarger{\sum}\limits_{i=1}^{N}}\Big(\mu_i(s) \times q[i,a_i]\Big)
	\label{approximate_Q_func}
\end{equation}
	The action-value function $Q(s,a)$ tells us how desirable it is to reach state $s$ by taking action $a$ by allowing to take the action $a$ many times and observe the return value.

	\item Calculate reward value: The controller receives the current values of $vm$ and $rt$ that correspond to the current state of the system $s$. The reward value $r$ is calculated based on two criteria: (i) the amount of resources acquired, which directly determine the cost, and (ii) SLO violations.

	\item Calculate the value of new state $s'$: By taking action $a$ and leave the system from the current state $s$ to the new state $s'$, the value of new state denoted $V(s')$ by is calculated by:
\begin{equation}
	V(s')= {\mathlarger\sum_{i=1}^{N}} \mu_i(s'). \underset{k}\max(q[i,a_k])
	\label{value_new_state}
\end{equation}
where $\max(q[i,a_k])$ is the maximum of the q-values applicable in the state $s'$.

\item Calculate error signal: As an \textit{on-policy} approach, SARSA estimates the value of action $a$ in state $s$ using experience actually gathered as it follows its policy, i.e., it always incorporates the actual agent's behavior. We mark $\Delta Q(s,a)$ as the error signal given by:
\begin{equation}
\Delta Q_{\texttt{FSL}}(s,a) = r + \gamma \times Q(s',a') - Q(s,a)
\label{delta_SARSA}
\end{equation}
where $\gamma$ is a discount rate which determines the relative importance of future rewards. A low value for $\gamma$ means that we value rewards that are close to time $t$, and a higher discount gives more value to the ones that are further in the future than those closer in time. 

	\item Update q-values: at each step, q-values are updated by :
\begin{equation}
	q[i,a_i]=q[i,a_i] + \eta.\Delta Q.\mu_i\big(s(t)\big)
\label{update_q_values}
\end{equation}	
where $\eta$ is the learning rate and takes a value between $0$ and $1$. Lower values for $\eta$ mean that preferring old values slightly with every update and a higher $\eta$ gives more impact on recent rewards.
	
\end{enumerate}

The \texttt{FSL} solution is sketched in Algorithm \ref{FL_SARSA}. 
\begin{algorithm}
\fontsize{10pt}{11pt}\selectfont
\caption{Fuzzy SARSA learning(\texttt{FSL})}
\label{FL_SARSA}
\begin{algorithmic}[1]
\Require discount rate ($\gamma$) and learning rate ($\eta$)

\State initialize q-values
\State observe the current state $s$
\State choose partial action $a_i$ from state $s$ ($\epsilon$-greedy strategy)
\State compute action $a$  from $a_i$ (Eq. \ref{output_eq}) and its corresponding quality $Q(s,a)$ (Eq. \ref{approximate_Q_func})
\Repeat
\State apply the action $a$, observe the new state $s'$ 
\State receive the reinforcement signal (reward) $r$ 

\State choose partial action $a'_i$ from state $s'$
\State compute action $a'$  from $a'_i$ (Eq. \ref{output_eq}) and its corresponding quality $Q(s',a')$ (Eq. \ref{approximate_Q_func})

\State compute the error signal $\Delta Q_{\texttt{FSL}}(s,a)$ (Eq. \ref{delta_SARSA})
\State update $q$-values (Eq. \ref{update_q_values})
\State $s \gets s'$ ,$a \gets a'$ 
\Until{convergence is achieved}
\end{algorithmic}
\end{algorithm}

\subsection{Fuzzy Q-Learning (\texttt{FQL})}

As we explained before, the major difference between Q-learning and the SARSA approach is their strategy to update $q$-values, i.e., in Q-learning $q$-values are updated using the largest possible reward (or reinforcement signal) from the next state. In simpler words, Q-learning is an \textit{off-policy} algorithm and updates $Q$-table values independent of the policy the agent currently follows. In contrast, SARSA as an \textit{on-policy} approach always incorporates the actual agent's behavior. Thus, the error signal for \texttt{FQL} is given by :
\begin{equation}
\Delta Q_{\texttt{FQL}}(s,a) = r + \gamma \times V(s') - Q(s,a)
\label{delta_Q_learning}
\end{equation}
The \texttt{FQL} is presented in  Algorithm \ref{FL_QL}

\begin{algorithm}
\fontsize{10pt}{11pt}\selectfont
\caption{Fuzzy Q-Learning (\texttt{FQL})}
\label{FL_QL}
\begin{algorithmic}[1]
\Require discount rate ($\gamma$) and learning rate ($\eta$)

\State initialize q-values
\State observe the current state $s$
\Repeat
\State choose partial action $a_i$ from state $s$ ($\epsilon$-greedy strategy)
\State compute action $a$  from $a_i$ (Eq. \ref{output_eq}) and its corresponding quality $Q(s,a)$ (Eq. \ref{approximate_Q_func})
\State apply the action $a$, observe the new state $s'$ 
\State receive the reinforcement signal (reward) $r$ 

\State compute the error signal $\Delta Q_{\texttt{FQL}}(s,a)$ (Eq. \ref{delta_Q_learning})
\State Update $q$-values (Eq. \ref{update_q_values})
\State $s \gets s'$ 
\Until{convergence is achieved}
\end{algorithmic}
\end{algorithm}

As an example, we assume the state space to be finite (e.g., 9 states as the full combination of $3 \times 3$ membership functions for fuzzy variables $w$ (workload) and $rt$ (response time). Our controller might have to choose a scaling action among 5 possible actions $\{\minus 2,\minus 1,  0, \plus 1, \plus 2\}$. However, the design methodology that we demonstrated in this section is general and can be applied for any possible state and action spaces. Note, that the convergence is detected when the change in the consequent functions is negligible in each learning loop.

\section{Implementation}\label{sec_5}

We implemented prototypes of the \texttt{FQL} and \texttt{FSL} algorithms in OpenStack. Orchestration and automation within OpenStack is handled by the Heat component. The auto-scaling decisions made by Heat on when to scale application and whether scale up/down should be applied, are determined based on collected metering parameters from the platform. Collecting measurement parameters within OpenStack is handled by Ceilometer (see Figure \ref{heat_Ceilometer}). The main part of Heat is the stack, which contains resources such as compute instances, floating IPs, volumes, security groups or users, and the relationship between these resources. Auto-scaling in Heat is done using three main resources: (i) \textit{auto-scaling group} is used to encapsulate the resource that we wish to scale, and some properties related to the scale process; (ii) \textit{scaling policy} is used to define the effect a scale process will have on the scaled resource; and  (iii) an \textit{alarm} is used to define under which conditions the scaling policy should be triggered.

In our implementation, the environment contains one or more VM instances that are controlled by a load balancer and defined as members in autoscaling group resources. Each instance (VM) includes a simple web server to run inside of it after launching. Each web server listens to an input port (here port 80), returns a simple HTML page as the response. User data is the mechanism by which users can define their own pre-configuration as a shell script (the code of web server) that the instance runs on boot.

\begin{figure}[H]
\center
	\scalebox{0.90}{\includegraphics[trim=50 495 320 165,clip=true]{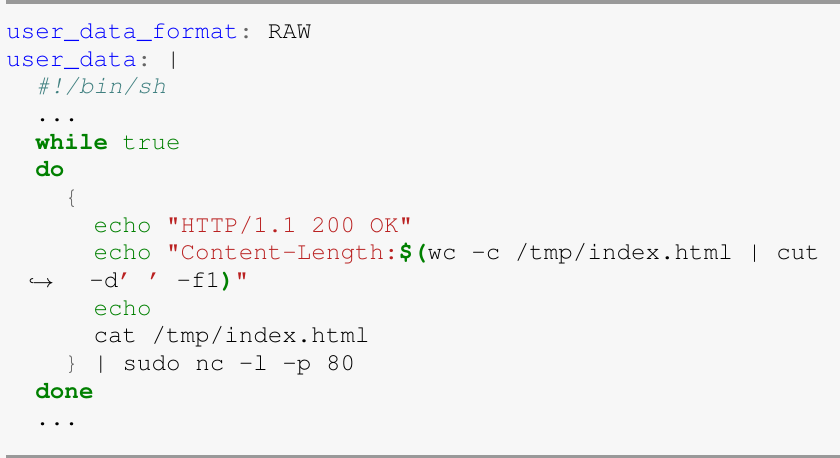}}
\caption{The simple web server}
\label{simple_web_server}
\end{figure}

In Fig.\ \ref{simple_web_server}, the template used for the web server is shown. For the VM web-server instance type, we used a minimal Linux distribution: the \texttt{cirros}\footnote{CirrOS images, https://download.cirros-cloud.net/} image was specifically designed for use as a test image on clouds such as OpenStack \cite{Arabnejad2016}.

The next step is defining the \textit{scaling policy}, which is used to define the effect a scaling process will have on the scaled resource, such as "add -1 capacity" or "add +10\% capacity" or "set 5 capacity". Figure \ref{scale_policy} shows the template used for the scaling policy.

\begin{figure}[H]
\vspace*{-0.5cm}
\center
	\scalebox{0.90}{\includegraphics[trim=50 180 320 525,clip=true]{./figs/code.pdf}}

\caption{The template for scaling policy}
\label{scale_policy}
\end{figure}

The scaling policy resource is defined as a type of \texttt{OS::Heat::ScalingPolicy} and its properties are as follows: 1) \texttt{auto\_scaling\_group\_id} is the specific scaling group ID to apply the corresponding scale policy, 2) \texttt{adjustment\_type} is the type of adjustment (absolute or percentage) and can be set to allowed values such as \textit{change in capacity}, \textit{exact capacity}, \textit{percent change in capacity}, and 3) \texttt{scaling\_adjustment} is the size of the adjustment in absolute value or percentage.

\begin{figure}[H]
\centering
	\scalebox{0.30}{\includegraphics[trim=160 10 140 60,clip=true]{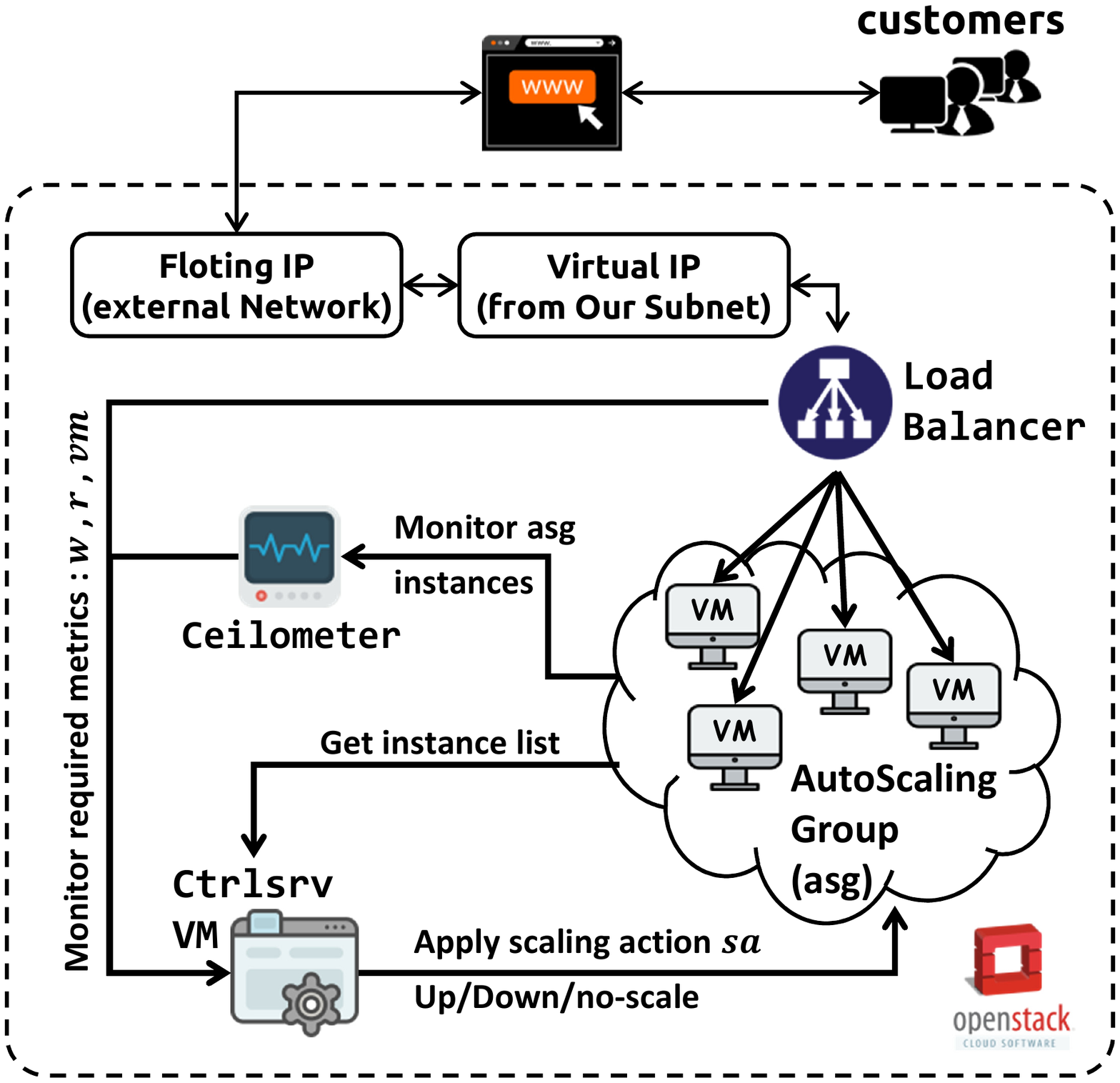}}
	\caption{Overview of the implemented system}
	\label{loadbalancer}
\vspace*{-0.7cm}
\end{figure}

We used our auto-scaling manager instead of the native auto-scaling tool in OpenStack, which is designed by setting alarms based on threshold evaluations for a collection of metrics from Ceilometer. For this threshold approach, we can define actions to take if the state of the watched resource satisfies specified conditions. However, we replaced this default component by the \texttt{FRL} approaches, to control and manage scaling options. In order to control and manage scaling option by the two \texttt{FRL} approaches (\texttt{FQL} and \texttt{FSL}), we added an additional VM resource, namely \texttt{ctrlsrv}, which acts as an auto-scaling server and enacts the scale up/down decision proposed by either of the two \texttt{FRL} approaches. For \texttt{ctrlsrv}, due to the impossibility of installing any additional package in the \texttt{cirros} image, we considered a VM machine running a Linux Ubuntu precise server. Figure \ref{loadbalancer} illustrates the implemented system in OpenStack. The created load balancer distributes client's HTTP request across a set of web-servers, i.e., auto-scaling group members, collected in load balancer pool. The algorithm used to distribute load between the members of the pool is \texttt{ROUND\_ROBIN}. 

Figure \ref{loadbalancer} shows the complete process of how the proposed fuzzy auto-scaling approach works. First, \texttt{ctrlsrv} gathers information from the  load balancer, ceilometer and the current state of members (web-servers) in an autoscaling group, then decides which horizontal scaling, i.e., up or down, should be applied to the target platform. For instance, the scale-up even will launch a new web-server instance, which may take a few minutes as the instance needs to be started, and adds it to the load-balancer pool. The two proposed auto-scaling algorithms, \texttt{FQL} and \texttt{FSL}, are coded and run inside of the \texttt{ctrlsrv} machine. We implemented and added a complete fuzzy logic library. This is functionally similar to the respective matlab features and implements our \texttt{FRL} approaches.

\begin{figure}[h] \centering
	\subfigure{\scalebox{0.20}{\includegraphics[trim=10 70 10 70,clip=true]{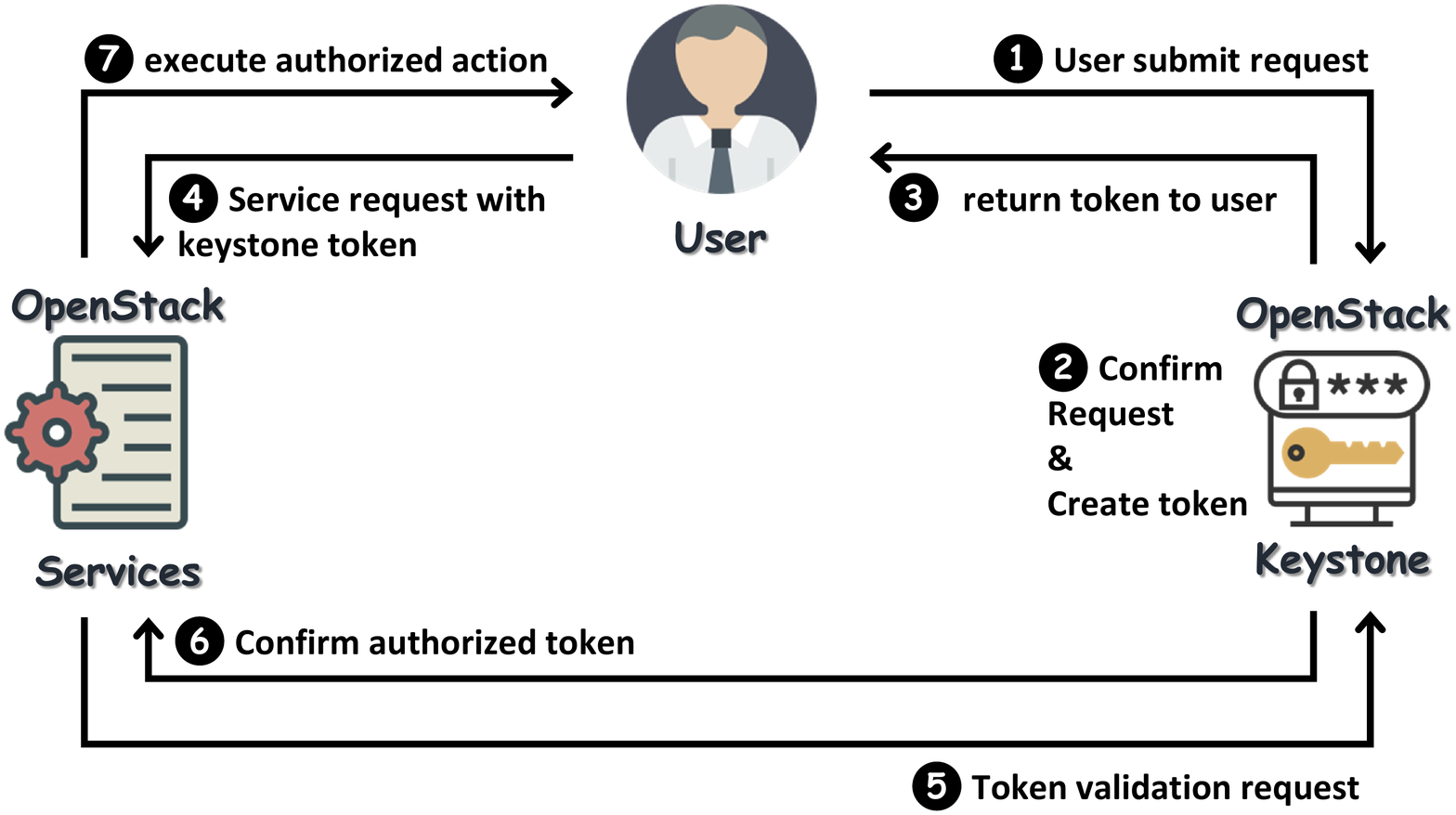}}}
	\caption{\texttt{cURL} process of calling OpenStack API }
	\label{curl}
\end{figure}

For some parameters in the proposed algorithm, such as the current number of VM instances or workload, we need to call the OpenStack API. For example, the command \texttt{nova list}  shows a list of running instances. The API is a RESTful interface, which allows us to send URL requests to the service manager to execute commands. Due to the unavailability of direct access to the OpenStack API inside of the \texttt{ctrlsrv} machine, we used the popular command line utility \texttt{cURL} to interact with a couple of OpenStack APIs. \texttt{cURL} lets us transmit and receive HTTP requests and responses from the command line or a shell script, which enabled us to work with the OpenStack API directly. In order to use an OpenStack service, we needed authentication. For some OpenStack APIs, it is necessary to send additional data, like the authentication key, in a header request. In Figure \ref{curl}, the process of using \texttt{cURL} to call OpenStack APIs is shown. The first step is to send a request authentication token by passing credentials (username and password) from OpenStack Identity service. After receiving \texttt{Auth-Token} from the Keystone component, the user can combine the authentication token and Computing Service API Endpoint to send an HTTP request and receive the output. We used this process inside the \texttt{ctrlsrv} machine to execute OpenStack APIs and collect required outputs.

By combining these settings, we are able to run both \texttt{FRL} approaches, i.e., \texttt{FQL} and \texttt{FSL}, as the manager and controller of auto-scaling process in OpenStack.

\section{Experimental Comparison}\label{sec_6}

The experimental evaluation aims to show the effectiveness of two proposed approaches  \texttt{FQL} and \texttt{FSL}, but also to look at differences. Furthermore, the cost improvement by proposed approaches for cloud provider is demonstrated.

\begin{figure*}[t!]
\begin{minipage}{1.00\textwidth}
\centering
	\subfigure[\scriptsize{Predictable Bursting pattern}]{\RTtrim{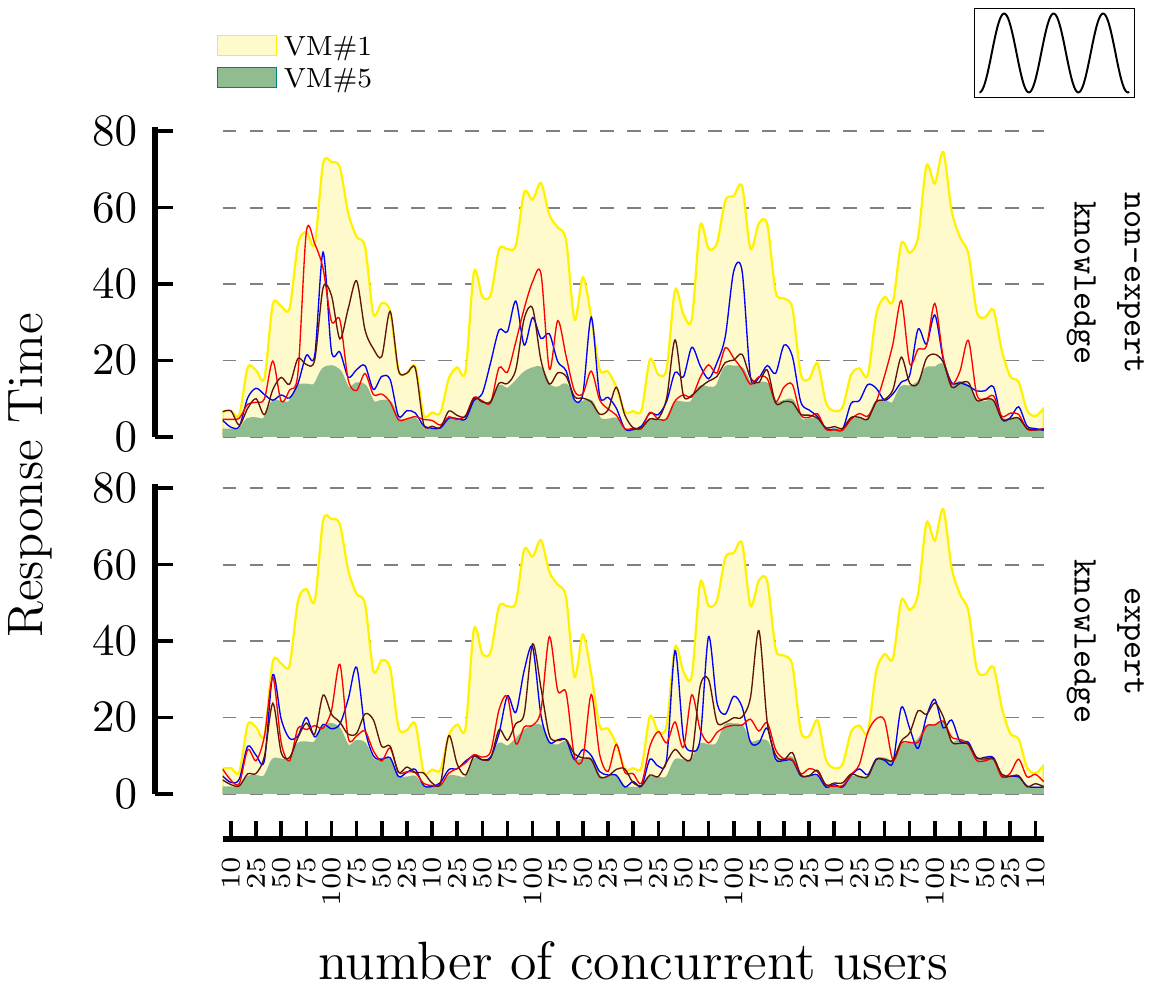}}
	\subfigure[\scriptsize{variations pattern}]{\RTtrim{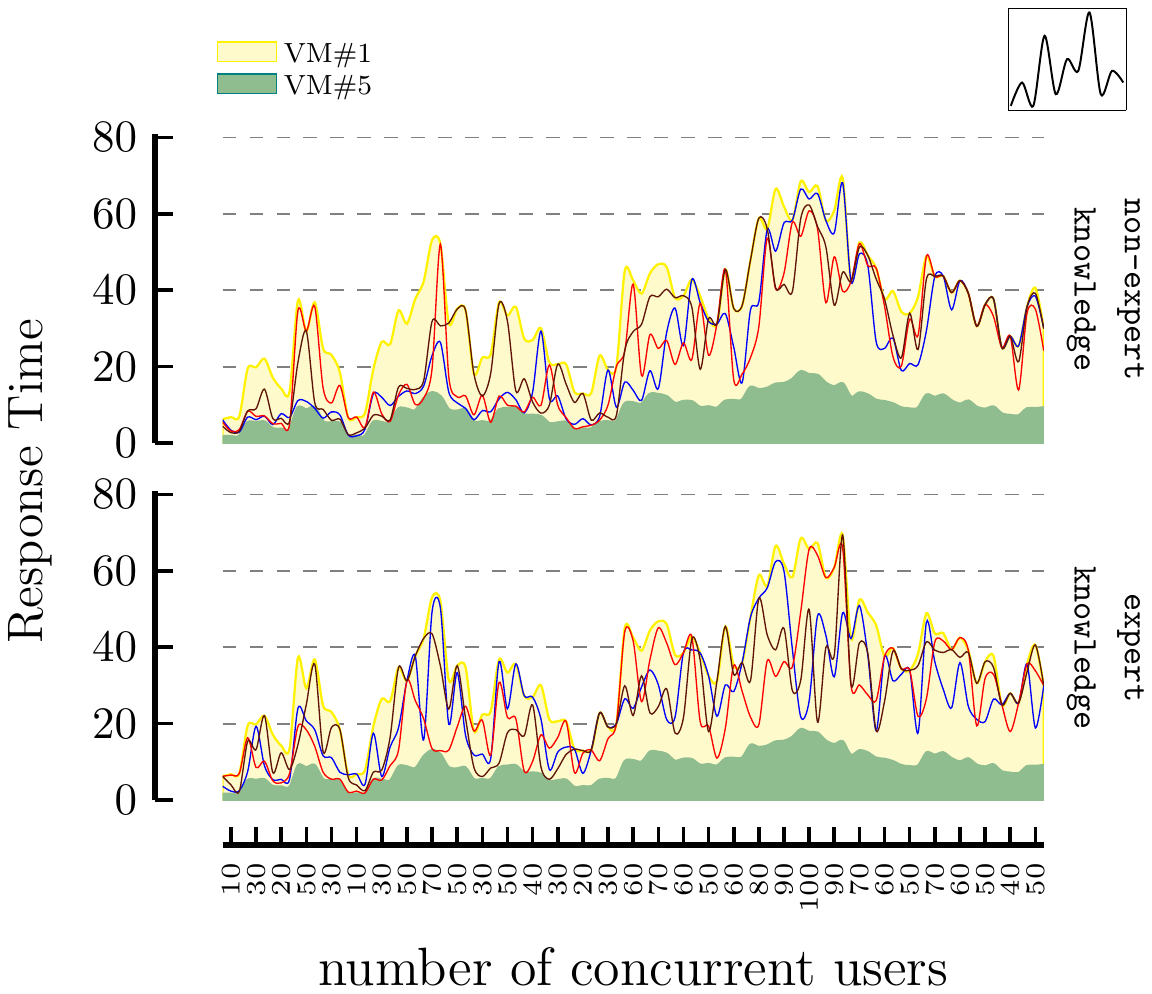}}
	\subfigure[\scriptsize{ON\&OFF pattern}]{\RTtrim{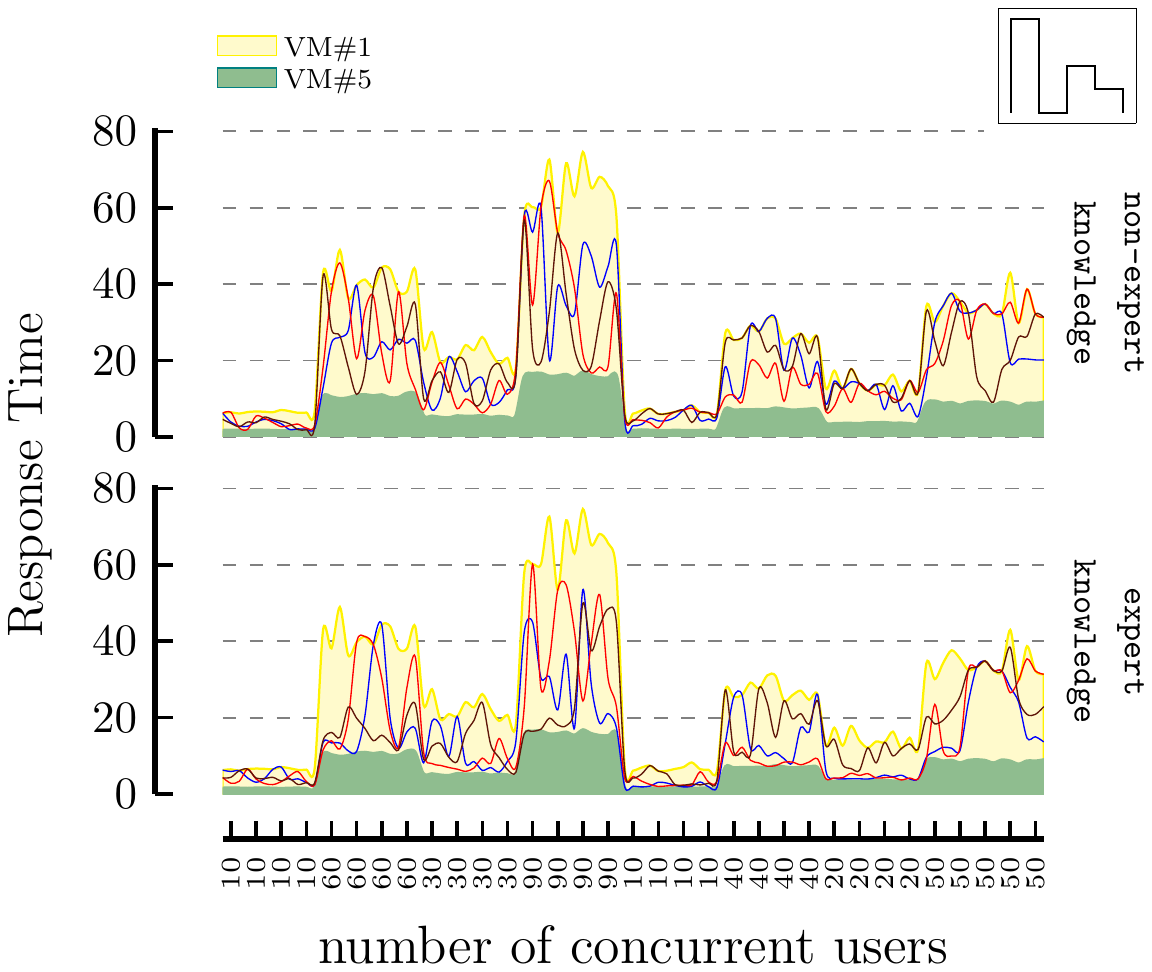}}
	\caption{The observed end-to-end response time of \texttt{FSL}}	
	\label{RT_FSL_fig}
\end{minipage}
\begin{minipage}{1.00\textwidth}
\centering
	\subfigure[\scriptsize{Predictable Bursting pattern}]{\RTtrim{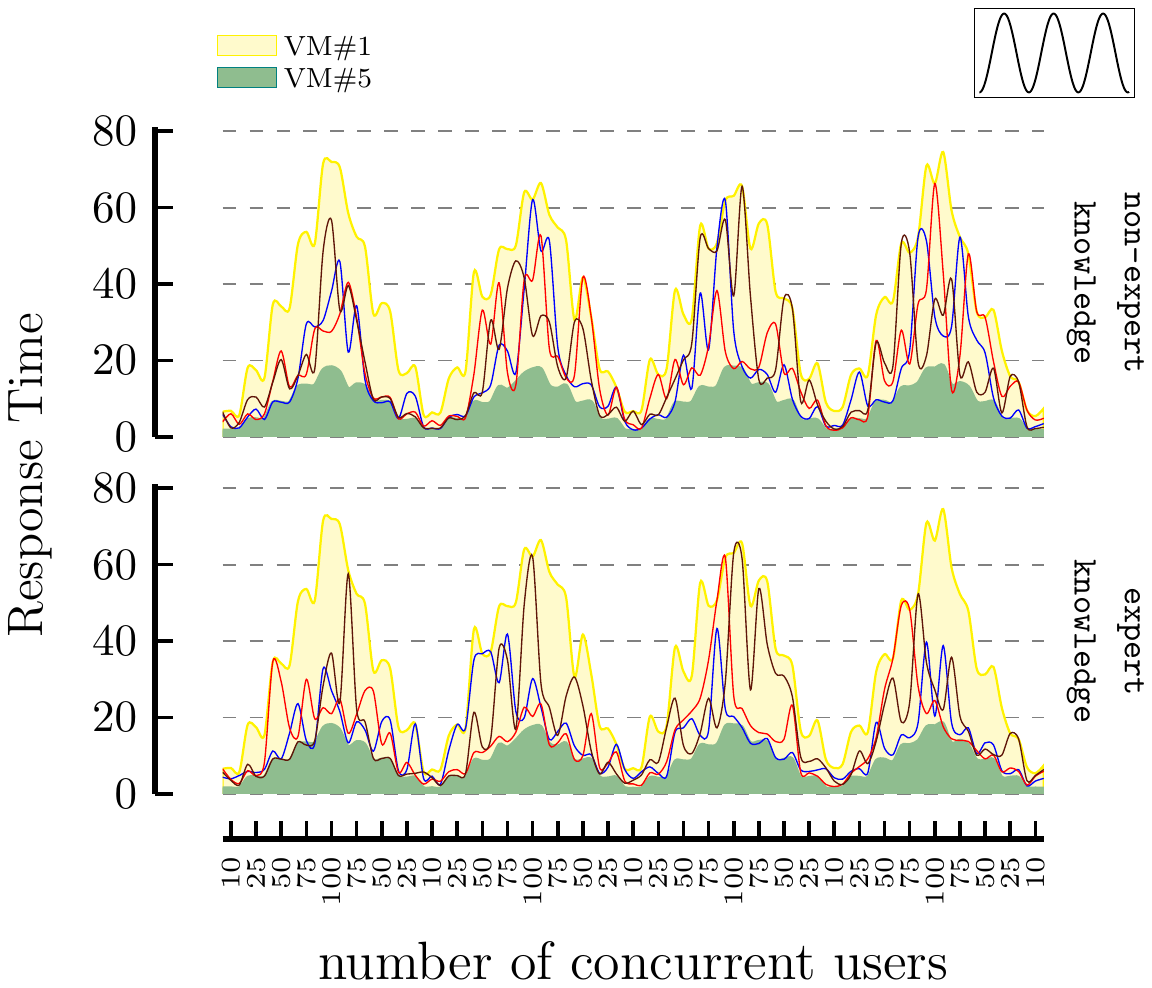}}
	\subfigure[\scriptsize{variations pattern}]{\RTtrim{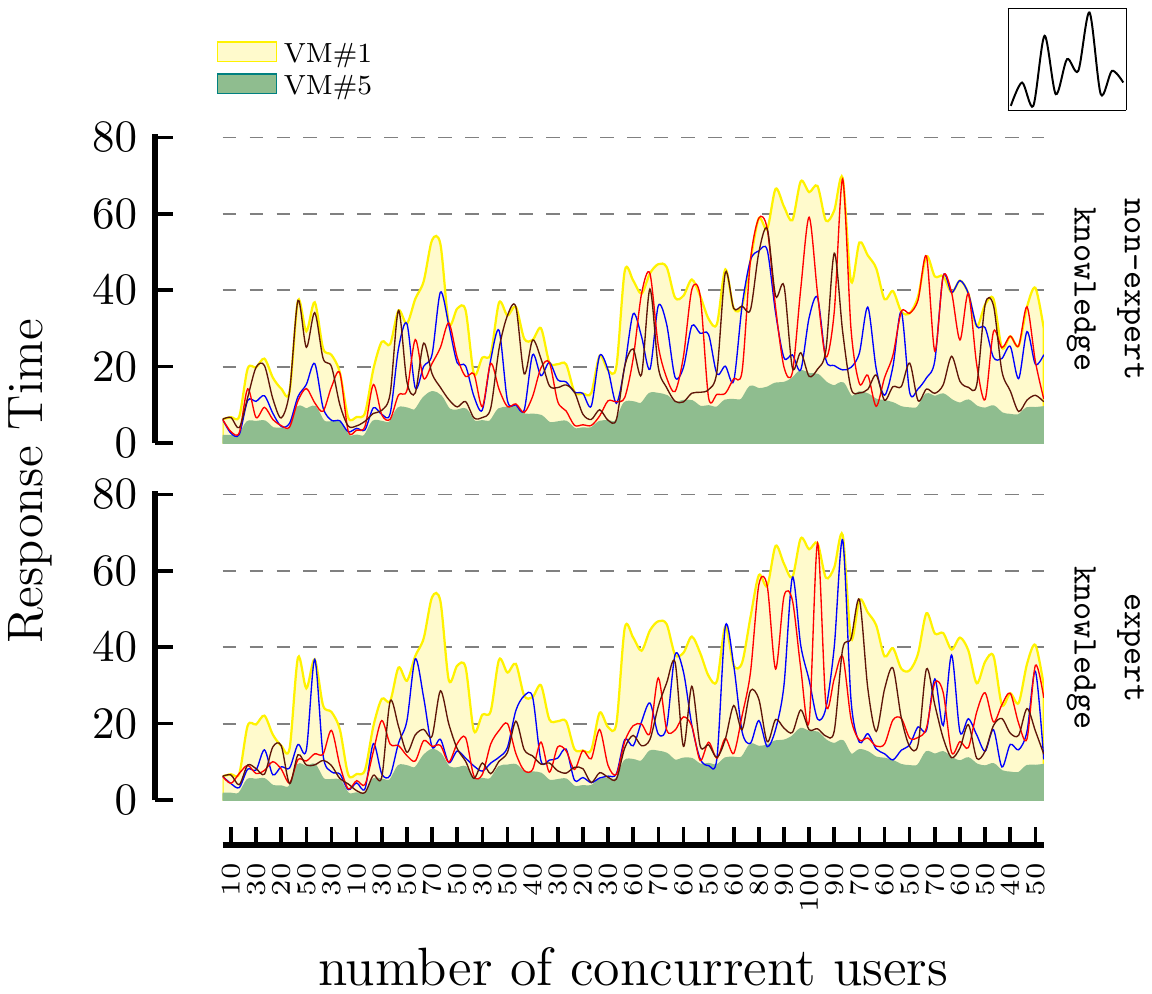}}
	\subfigure[\scriptsize{ON\&OFF pattern}]{\RTtrim{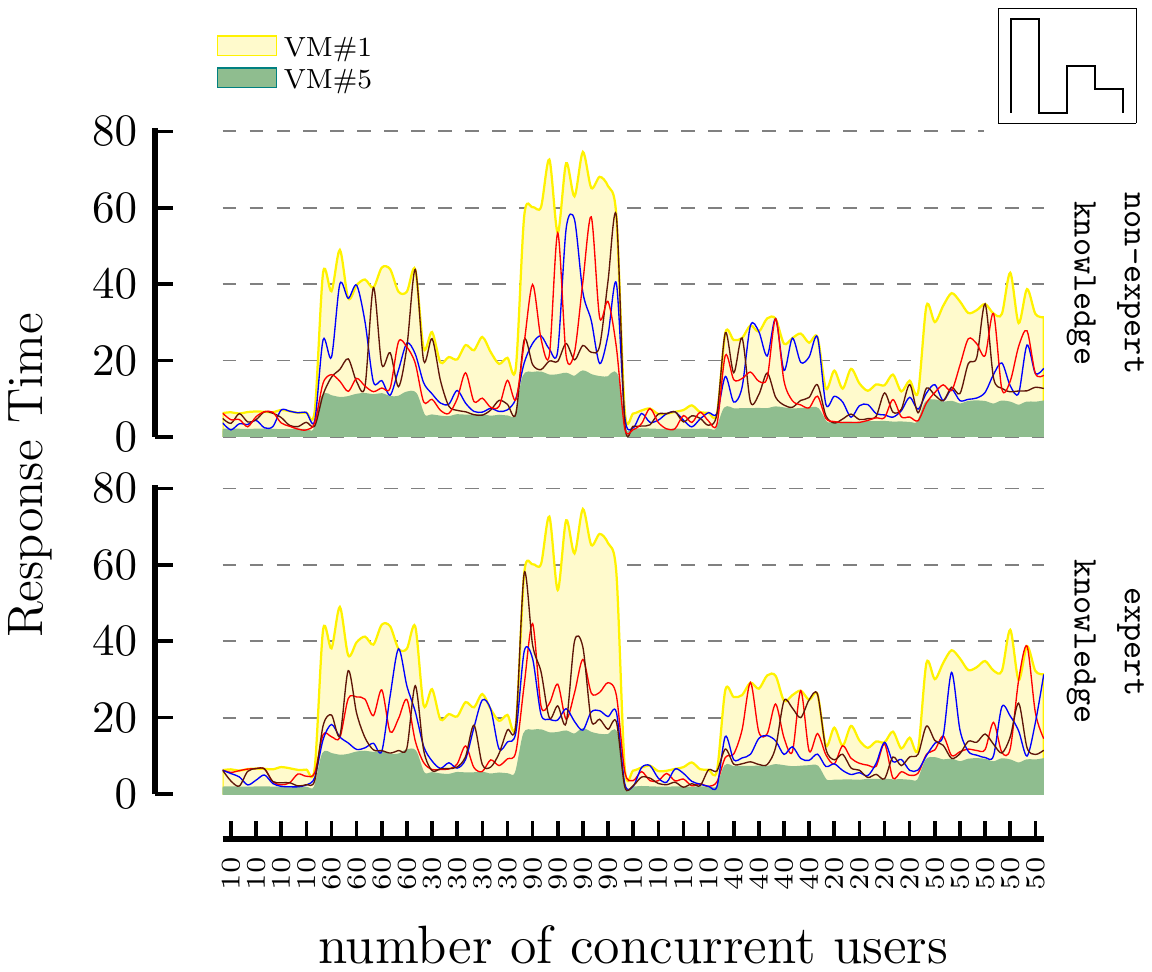}}
	\caption{The observed end-to-end response time of \texttt{FQL}}	
	\label{RT_FQL_fig}
\end{minipage}
\end{figure*}

\begin{figure*}[t!]
\begin{minipage}{0.50\textwidth}
\centering
	\subfigure[\scriptsize{Predictable  Bursting \newline pattern}]{\VMusagetrim{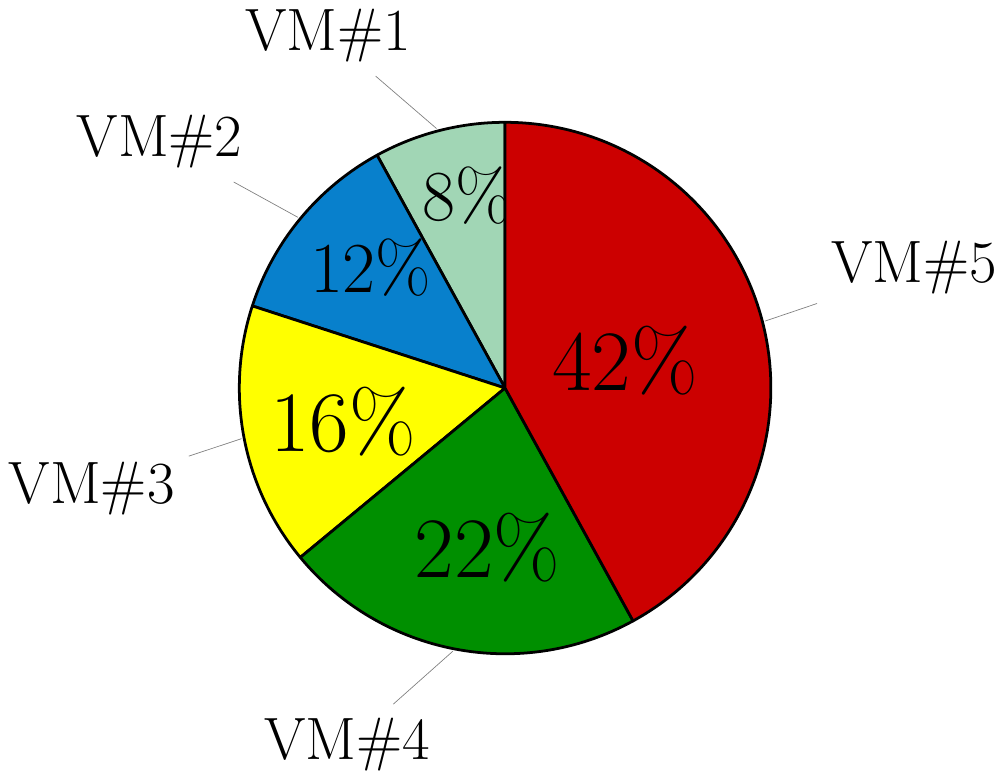}}
	\subfigure[\scriptsize{variations  pattern}]{\VMusagetrim{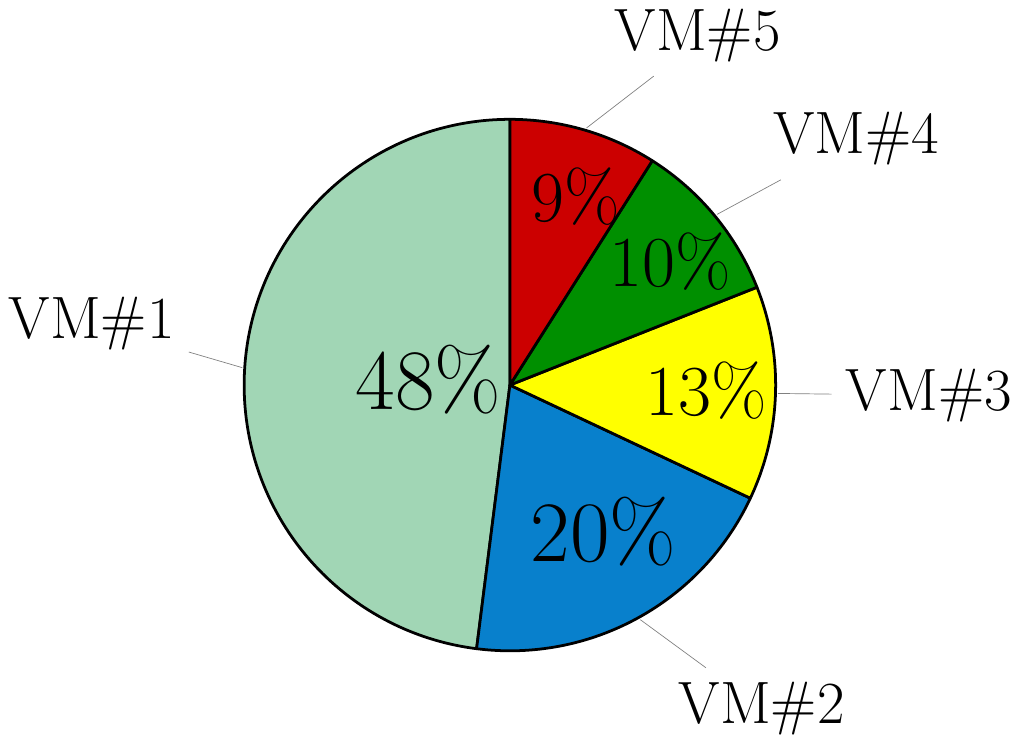}}
	\subfigure[\scriptsize{ON\&OFF pattern}]{\VMusagetrim{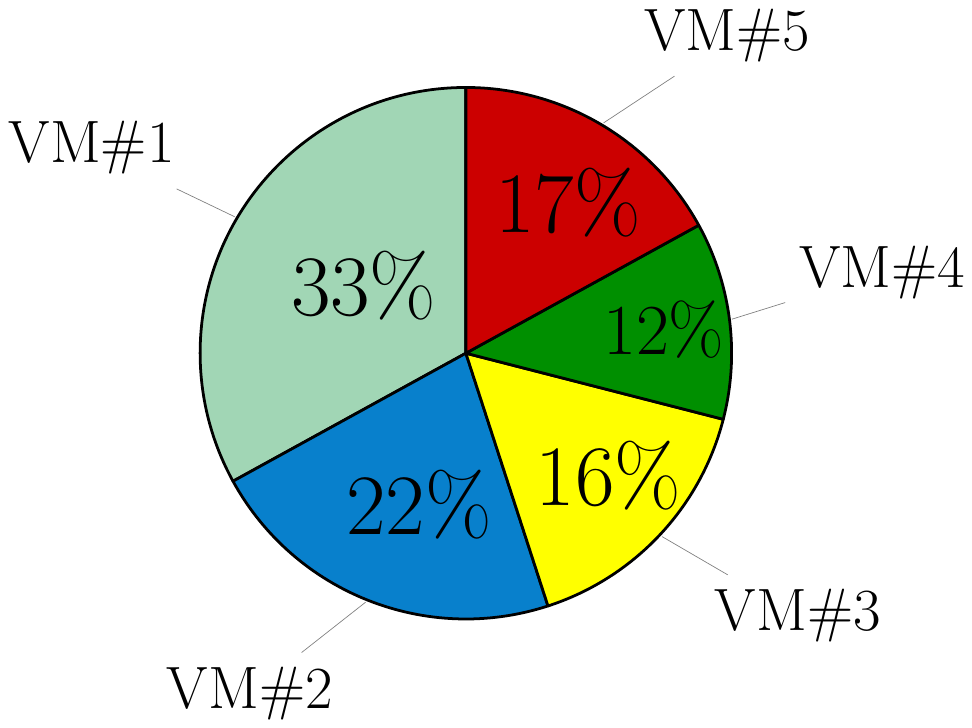}}
	\caption{Percentage number of VMs used by \texttt{FSL}}
	\label{VM_usage_FSL_fig}
\end{minipage}
\begin{minipage}{0.50\textwidth}
\centering
	\subfigure[\scriptsize{Predictable Bursting \newline pattern}]{\VMusagetrim{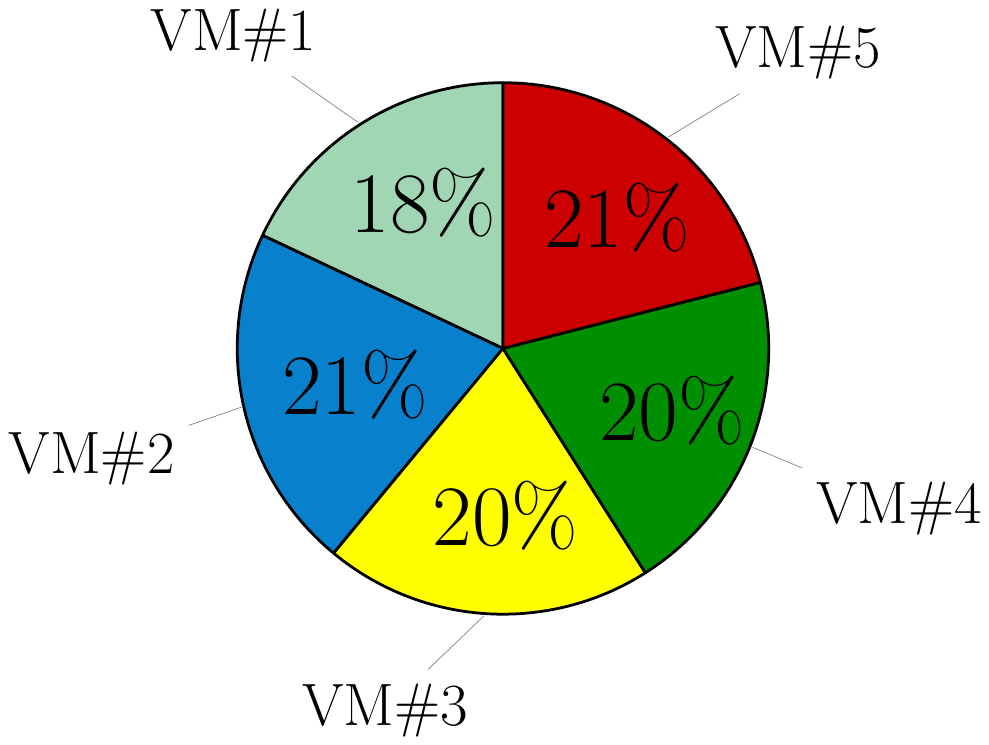}}
	\subfigure[\scriptsize{variations pattern}]{\VMusagetrim{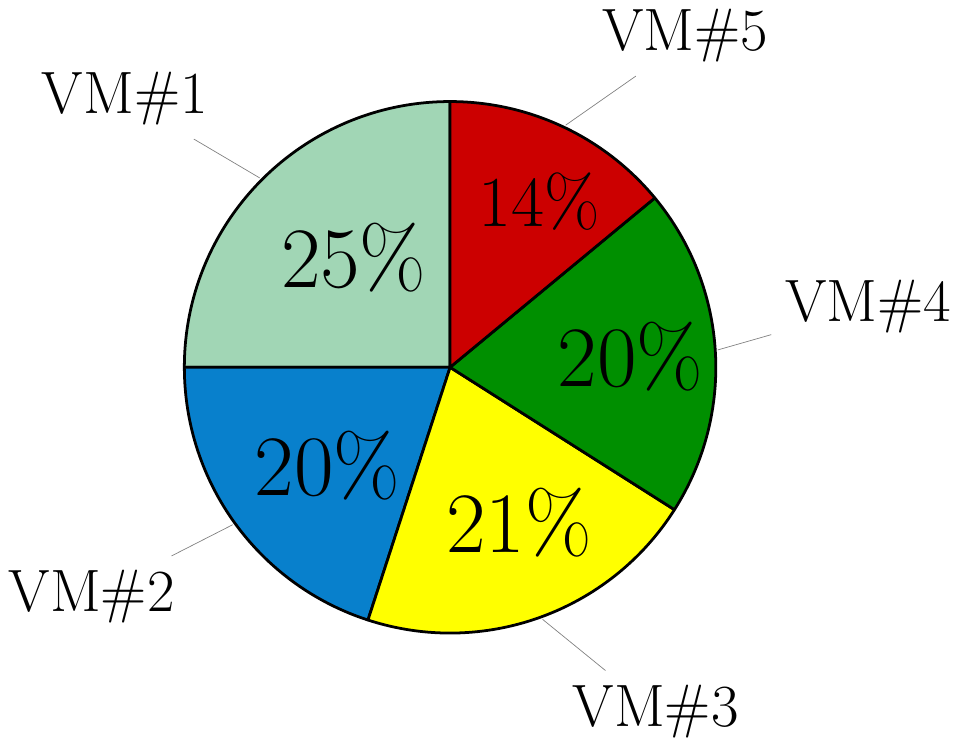}}
	\subfigure[\scriptsize{ON\&OFF pattern}]{\VMusagetrim{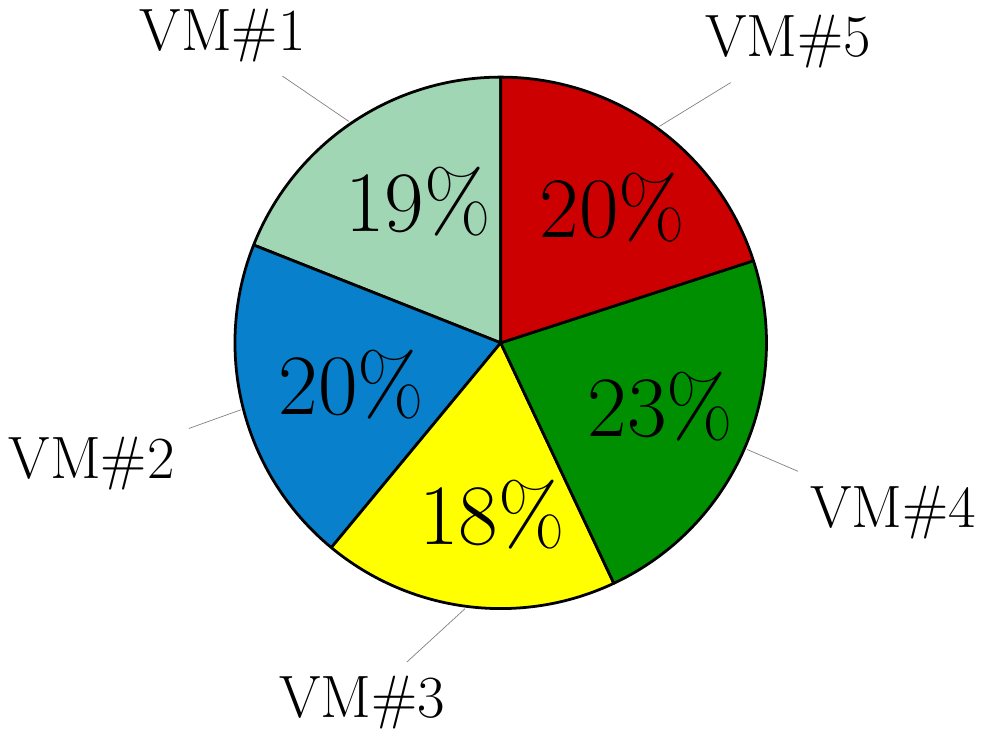}}
	\caption{Percentage number of VMs used by  \texttt{FQL}}
	\label{VM_usage_FQL_fig}
\end{minipage}
\end{figure*}

\begin{figure}
\vspace*{-0.1cm}
\centering
	\scalebox{0.60}{\includegraphics[trim=30 190 480 130,clip=true]{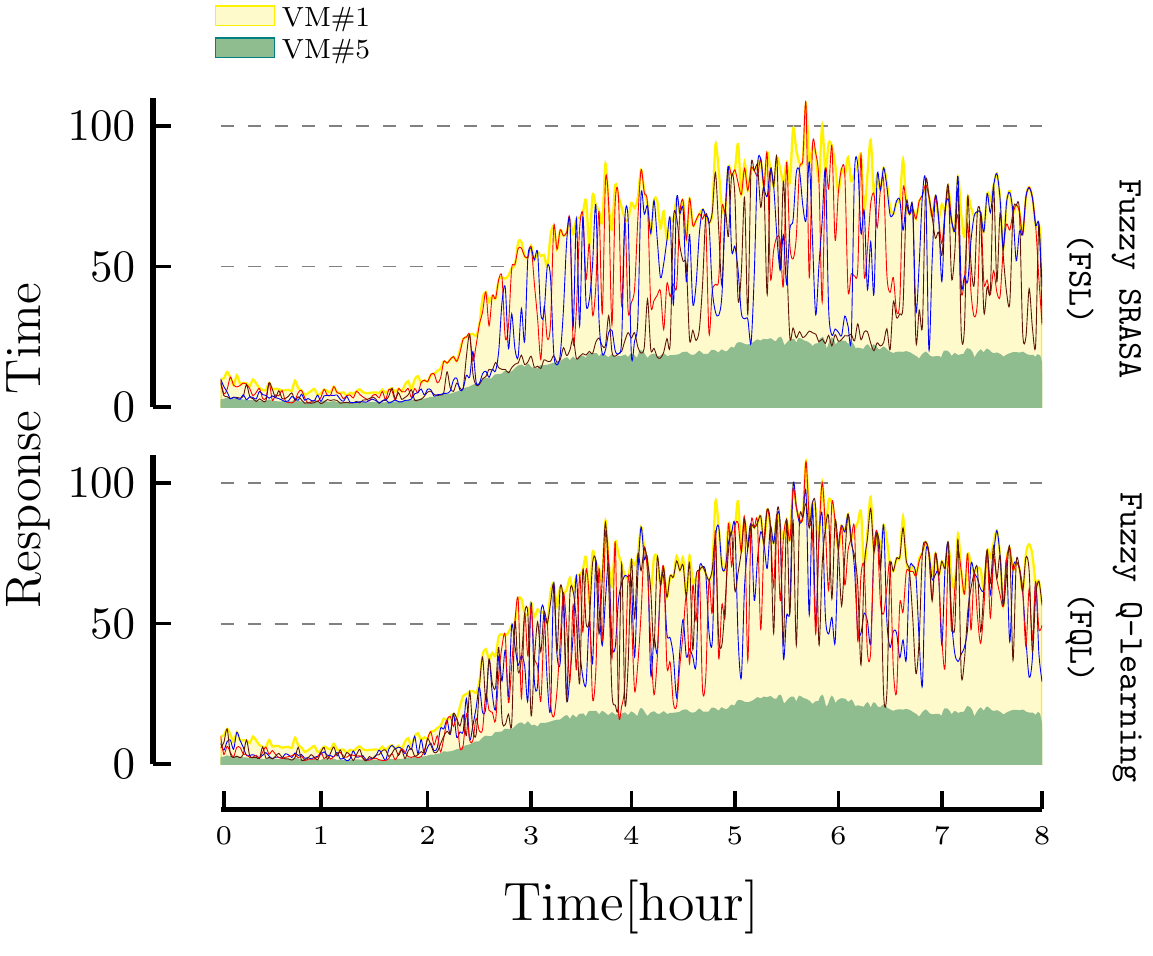}}
	\vspace*{-0.3cm}
	\caption{The observed end-to-end response time for Wikipedia workload}
	\label{RT_wiki}
\end{figure}

\begin{figure}
\vspace*{-0.3cm}
\centering
	\scalebox{0.60}{\includegraphics[trim=30 190 480 130,clip=true]{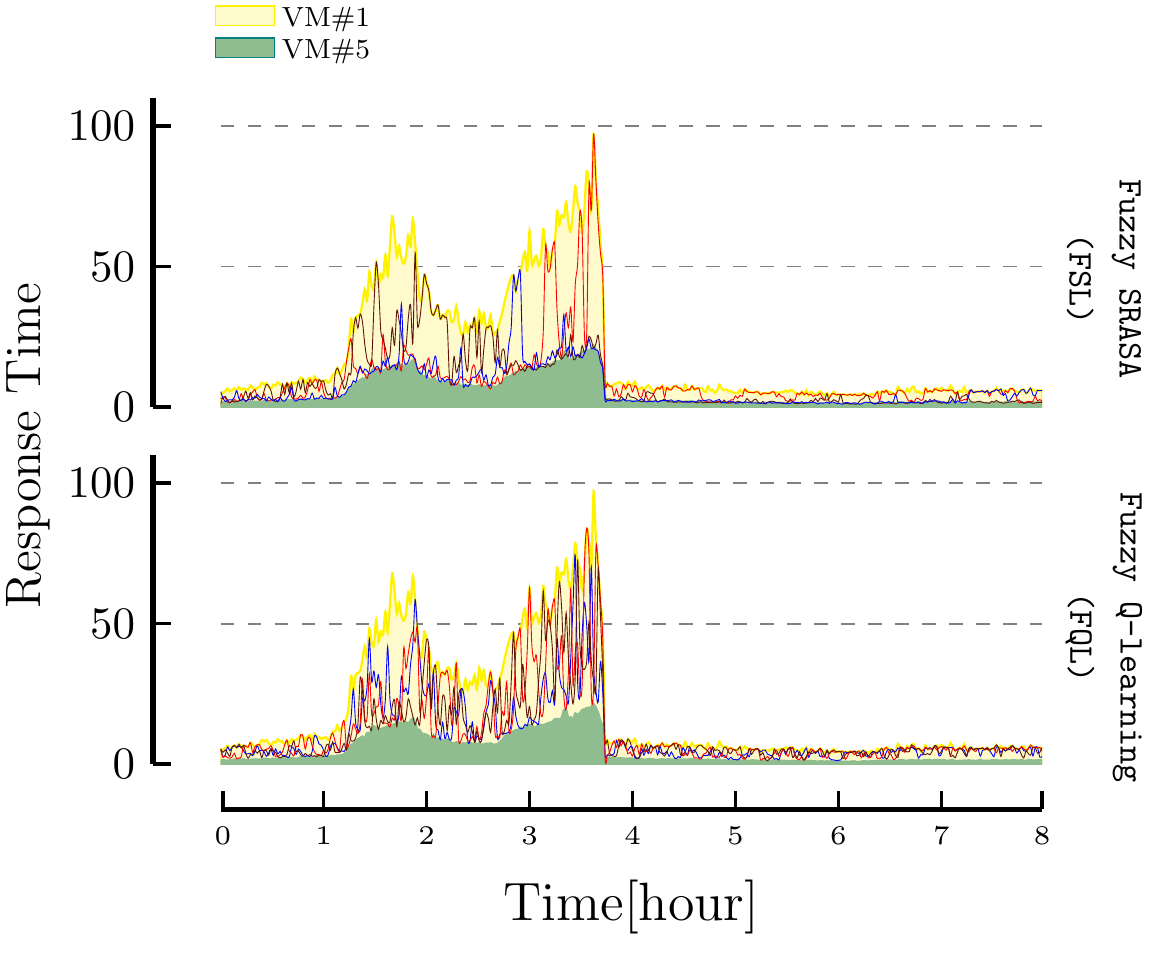}}
	\vspace*{-0.3cm}
	\caption{The observed end-to-end response time for FIFA'98 workload}
	\label{RT_FIFA}
\end{figure}

\subsection{Experimental setup and benchmark}

In our experiment, the two proposed approaches \texttt{FQL} and \texttt{FSL} were implemented as full working systems and were tested in the OpenStack platform. As the required parameters, the maximum and minimum number of VMs that were allowed to be available at same time were set to $5$ and $1$, respectively. Here, we considered low number of VMs to demonstrate the effectiveness of our proposed approaches under heavy load user request traffic. However, larger VM number can be applied for these parameters. The term workload refers to the number of concurrent user request arrivals in given time. Workload is defined as the sequence of users accessing  the target application that needs to be handled by the auto-scaler. 

Application workload patterns can be categorized in three representative patterns \cite{lorido2014review}: (a) the \textit{Predictable Bursting} pattern indicates the type of workload that is subject to periodic peaks and valleys typical for services with seasonality trends or high performance computing, (b) the \textit{Variations} pattern reflects applications such as News\&Media, event registration or rapid fire sales, and (c) the \textit{ON\&OFF} pattern reflects applications such as analytics, bank/tax agencies and test environments. In all cases, we considered $10$ and $100$ as minimum and maximum number of concurrent users per second. 

Additionally, we validated our approaches with real user request traces of the Wikipedia\footnote{Wikipedia Access Traces : http://www.wikibench.eu/?page\_id=60} and the FIFA WorldCup\footnote{FIFA98 View Statistics : http://ita.ee.lbl.gov/html/contrib/WorldCup.html} websites, which are the number of requests/users accessing these two websites per unit time. We used \texttt{Siege}\footnote{https://www.joedog.org/siege-home/}, a HTTP load testing and benchmarking utility, as our performance measuring tools. It can generate concurrent user requests, and measure the performance metric such as average response time. For each concurrent user number $N$ , we generate $N$ requests per second by \texttt{Siege} for $10$ minutes.

For fuzzy controller parameters, the learning rate is set to a constant value $\eta=0.1$ and the discount factor is set to $\gamma = 0.8$. Here, we considered lower value for $\eta$, thus giving more impact on old rewards with every update. After sufficient epochs of learning, we decrease the exploration rate ($\epsilon$) until a minimum value is reached, which here is $0.2$. \texttt{FRL} approaches start with an exploration phase and after the first learning convergence occurs, they enter the balanced exploration-exploitation phase. 

Additionally, we compared the two proposed approaches with a base-line strategy. The results of comparing with fixed numbers of VMs equal to a minimum and maximum permitted value are also shown as based-line (benchmark) approaches, named $VM\#1$ and $VM\#5$, reflecting under- and over-provisioning strategies.

Furthermore, in order to investigate the effects of initialized knowledge, we considered two types of fuzzy inference system (FIS) as the primary knowledge for fuzzy controller, \textit{expert} and \textit{not-expert} knowledge.

\subsection{Comparison of effectiveness}

Figures \ref{RT_FSL_fig} and \ref{RT_FQL_fig} show the fluctuation of the observed end-to-end response time for three type of workload patterns obtained by two approaches \texttt{FSL} and \texttt{FQL}, respectively. In order to investigate the behaviour of the auto-scaler, we considered two types of initialized knowledge (expert and non-expert) and each algorithm \texttt{FQL} and \texttt{FSL} was executed several times and represented by a different color in presented figures.

During the test, workloads were dynamically changed. Depending on incoming workload (the concurrent input requests submitted by individual users) and the number of available VMs, corresponding response timed varied between upper or lower bound. Both \texttt{FQL} and \texttt{FSL} algorithms with adaptive policies continuously monitored these fluctuation of the response time and identified workload changes. The scaling decisions were applied accordingly as recommended by the proposed algorithms. In our experiment, the up/down scaling process can be completed in a few seconds, due to simplicity and fast booting of \texttt{Cirros} image.

We compared  \texttt{FQL} and \texttt{FSL} with $VM\#1$ and $VM\#5$ as the base-line approaches, which have a fixed number of VMs during the test.
Figures \ref{RT_FSL_fig} and \ref{RT_FQL_fig} show that the proposed auto-scalers are able to dynamically set the number of required resources to the current workload, providing only resource allocations that are needed to meet the user's QoS expectations.  
As seen from Figures \ref{RT_FSL_fig} and \ref{RT_FQL_fig}, both algorithms \texttt{FQL} and \texttt{FSL} adapt themself to input workload in order to meet SLA parameters, which here is the response time. 

The difference in the algorithms can be seen from the quality of the solution, i.e., the scaling value. Both algorithms represent dynamic resource provisioning to satisfy upcoming resource demand. However:
\begin{enumerate}[leftmargin=*]
\item
For the \textit{Predictable Burst} workload pattern (Figures \ref{RT_FSL_fig}(a) and \ref{RT_FQL_fig}(a)),  \texttt{FSL} finds a significantly better solution compared to \texttt{FQL}. The reason can be explained by the speed of convergence for each RL approach. Q-learning does not learn the same policy as it follows which consequences that it learns slower. This means that although the learning improves the approximation of the optimal policy, it does not necessarily improve the policy which is actually being followed. 
On the other hand, \textit{on-policy} learning used in by \texttt{FSL} learns faster and enters the balanced exploration-exploitation phase, i.e., completes learning phase quickly and reaches a minimum exploration rate ($\epsilon$) that avoids more exploration in the action selection step.
\item
As a result of the performance improvement achieved by SARSA, \texttt{FSL} has a tendency to get more VMs launched to obtain a good solution which can be realized by comparing the percentage number of VMs used by these two algorithms (Figure \ref{VM_usage_FSL_fig}(a) and Figure \ref{VM_usage_FQL_fig}(a)). 
\item
For the \textit{Variations} workload pattern, \texttt{FQL} is superior to the solution found  by \texttt{FSL} approach. 
Due to faster learning of the \textit{on-policy} approach used in \texttt{FSL} alongside high fluctuation and non-periodic behaviour of this pattern, the non-explorative policy used after the learning phase is not optimized for the these workloads.
For the \textit{ON\&OFF} (Figures \ref{RT_FSL_fig}(c) and \ref{RT_FQL_fig}(c)) workload patterns, the value of the solution is more and less similar.
\end{enumerate}

The effectiveness of having expert (optimal) knowledge can be figured out by comparison between the two types of initial knowledge used for the experiment. In all presented cases, the good initial knowledge significantly improves the quality of results compared to non-expert (sub-optimal) knowledge. 

In addition, to validate the applicability of approaches against real-life situations, we used two real workloads: the Wikipedia workload and  the FIFA WorldCup Website access logs. While the Wikipedia workload shows a steady and predictable trend, the FIFA workload has a bursty and an unpredictable pattern. 
 For the Wikipedia trace in figure \ref{RT_wiki}, \texttt{FSL} shows slightly better performance compared to \texttt{FQL}. 
 For the FIFA results shown in Figure \ref{RT_FIFA}, the situation is different. \texttt{FSL} as an \textit{on-policy} approach behaves better in terms of the measured response time, while \texttt{FQL} is still in exploration/exploitation phase.

\subsection{Comparison of cost-effectiveness of scaling}\label{cost_FRL}

Figures \ref{VM_usage_FSL_fig} and \ref{VM_usage_FQL_fig} show percentage numbers of used VMs for all workload patterns. The  approaches work on the current workload and relative response time of the system at the current time, increasing the number of available VMs (scale-up) and decreasing the number of idle VMs (scale-down). Both \texttt{FQL} and \texttt{FSL} conduct distributed-case scaling and allocate suitable numbers of VMs according to the workload. 

For different types of workload patterns, the average maximum number of VMs used during our experiment by \texttt{FQL} and \texttt{FSL} algorithms are $18.3\%$ and $22.6\%$, respectively. This implies our approaches can meet the QoS requirements using a smaller amount of resources, which is an improvement on resource utilisation for applications in terms of hosting VMs. Thus, the \texttt{FQL} and \texttt{FSL} approaches can perform auto-scaling of application as well as save cloud provider cost by increasing resource utilisation.

\section{Conclusion}

We investigated horizontal scaling of cloud applications. Many commercial solutions use simple approaches such as threshold-based ones. However, providing good thresholds for auto-scaling is challenging. Recently, machine learning approaches have been used to complement and even replace expert knowledge to design self-adaptable solutions to capable to react to unpredictable workload fluctuations. 

We proposed a fuzzy rule-based system, based on which we compared two well-know RL approaches, resulting in Fuzzy Q-learning (\texttt{FQL}) and Fuzzy SARSA learning (\texttt{FSL}). Both approaches can efficiently scale up/down cloud resources to meet the given QoS requirements while reducing cloud provider costs by improving resource utilisation. However, differences also emerge. In the SARSA experiment,  given the reward at each time step improves the quality of solutions for periodic workload pattern. Both algorithms have been implemented in OpenStack, an open-source IaaS platform, to demonstrate the practical effectiveness of proposed approach  has been successfully tested and presented and the validity of the comparison results are established.

In conclusion, this paper identifies the promising auto-scaling concepts for cloud computing: (i) developing an autonomic and complete auto-scaler for a cloud platform system by combining of techniques such as a fuzzy logic system and reinforcement learning to provide optimal resource management approach tailored to different types of workload pattern, and (ii) defining the concept of a complex auto-scaler, that can replace traditional threshold-based ones, (iii) implement the proposed auto-scaler in an open-source cloud platform and presenting results for different type of workloads. 

We have demonstrated the overall suitability of the different types of on-policy and off-policy RL approaches for auto-scaling, but also differences for specific workload patterns and converging times.
We plan to extend our approach in a number of ways: (i) extending \texttt{FQL4KE} to perform in environments which are partially observable, (ii) exploiting clustering approaches to learn the membership functions of the antecedents (in this work we assume they do not change once they specified, for enabling the dynamic change we will consider incremental clustering approaches) in fuzzy rules and (iii) look at other resource types such as containers \cite{Pahl2015}.

\section{Acknowledgement}
This work was partly supported by IC4 (Irish Centre for Cloud Computing and Commerce), funded by EI and the IDA.

\bibliographystyle{plain}
\bibliography{ref}

\end{document}